\documentclass[12pt]{article}

\usepackage{amsmath}
\usepackage{epsfig}
\usepackage{amssymb,amsfonts,amsmath,amscd}
\usepackage[all]{xy}
\usepackage{pstricks}
\usepackage{pst-node}
\DeclareFontFamily{U}{rsf}{}
\DeclareFontShape{U}{rsf}{m}{n}{
  <5> <6> rsfs5 <7> <8> <9> rsfs7 <10-> rsfs10}{}
\DeclareMathAlphabet\Scr{U}{rsf}{m}{n}

\advance\voffset by -2.0cm
\advance\hoffset by -1.25cm
\def\Tr{\,{\rm Tr}\, }

\def\be{\begin{equation}}
\def\ee{\end{equation}}
\def\ba{\begin{eqnarray}}
\def\ea{\end{eqnarray}}

\def\cA{{\cal A}}
\def\e{{\rm e}}
\def\oW{\overline{W}}

\def\beq{\begin{equation}}
\def\eeq{\end{equation}}
\def\beqa{\begin{eqnarray}}
\def\eeqa{\end{eqnarray}}
\newcommand{\ZZ}{\mathbb{Z}}
\newcommand{\one}{{\bf 1}}
\newcommand{\cH}{{\cal H}}
\newcommand{\cM}{{\cal M}}
\newcommand{\cC}{{\cal C}}

\newcommand{\kket}[1]{\vert \Scr{B}, #1\rangle\!\rangle}
\newcommand{\bbra}[1]{\langle\!\langle \Scr{B},#1 \vert}
\newcommand{\cket}[1]{\vert \Scr{C},#1\rangle\!\rangle}
\newcommand{\cbra}[1]{\langle\!\langle \Scr{C},#1 \vert}
\newcommand{\half}{\frac{1}{2}}
\newcommand{\no}{\nonumber}
\newcommand{\nn}{\nonumber}
\newcommand{\Hom}{{\rm Hom}}
\renewcommand{\H}{{\cal H}}

\begin{document}

\vspace*{-1.5cm}
\thispagestyle{empty}
\begin{flushright}
hep-th/0612108
\end{flushright}
\vspace*{2.5cm}

\begin{center}
{\Large 
{\bf Permutation Orientifolds}}
\vspace{2.0cm}

{\large Ilka Brunner}%
\footnote{{\tt E-mail: brunner@itp.phys.ethz.ch}} 
{\large and} {\large Vladimir Mitev}%
\footnote{{\tt E-mail: vladimir.mitev@desy.de}} 
\vspace*{0.5cm}

Institut f\"ur Theoretische Physik, ETH-H\"onggerberg\\
8093 Z\"urich, Switzerland\\
\vspace*{2cm}

{\bf Abstract}
\end{center}
\noindent
We consider orientifold actions involving the permutation
of two identical factor theories. The corresponding crosscap states
are constructed in rational
conformal field theory. We study group manifolds, in particular the
examples $SU(2) \times SU(2)$ and $U(1)\times U(1)$ in detail, comparing
conformal field theory results with geometry.
We then consider orientifolds of tensor products of $N=2$ minimal models, 
which have a description as coset theories in
rational conformal field theory and as Landau Ginzburg models.
In the Landau Ginzburg language, B-orientifolds and D-branes are
described in terms of matrix factorizations of the superpotential.
We match the factorizations with the corresponding crosscap states.

\newpage


\section{Introduction}

Orientifolds \cite{orientifolds} play a prominent role in the context
of model building. For this and other reasons it is of interest to
understand their construction and physical properties in detail.

Given a tensor product of two identical theories, it is a natural
idea to consider orientifolds that involve the permutation of the
two factors. Such models can then be studied from various points
of view. If each single factor is a rational conformal field theory,
one can approach the problem algebraically and construct a crosscap
state describing a parity action that combines the permutation with
world sheet orientation reversal and possibly other actions. For a single
factor, this approach has been employed initially in \cite{PSS} 
 \footnote{A novel approach in the context of
rational conformal field theory is taken in \cite{FRS}, where a full
construction of the correlators on orientable and non-orientable
world sheets is given. We will not employ it in the current paper.}  .
On the other hand, the problem can also be approached geometrically. Given a sigma-model
with target space $M\times M$ one can consider an involution exchanging
the two factors, the fixed point set being the diagonally embedded 
$M\subset M\times M$.

In this paper, we will be interested in models that can be studied
from both points of view. 

We start out setting the stage in rational conformal field theory,
giving an explicit construction of permutation crosscap states. 
One motivation for the form of the crosscap state comes from the
idea that geometrically the  orientifold should be localized on the diagonal
of the two factors. The parity should act trivially on the corresponding
D-brane, whose world volume coincides with the orientifold fixed
point set. This is enough to determine the crosscap state by
a modular transformation to the closed string channel provided
that the boundary state is known. We also
point out that the form of our crosscap state suggests a
generalization of the arguments of \cite{GaGa,BFS} for
boundary states corresponding to automorphism twisted
boundary conditions to the orientifold
case.
Necessary
conditions for the consistency of our proposed crosscap states 
are provided by the one-loop diagrams,
which we summarize in the appendix. 

In section 3 we 
consider orientifolds of products of group manifolds $G\times G$. 
Even though these manifolds do not appear directly as part of a string model, they are simple models for studying the
behavior of D-branes and orientifold planes.
They also provide the basics for coset constructions such as those appearing
in Gepner models. 

The most symmetric D-branes preserving the full
underlying symmetry were constructed as rational boundary
states by Cardy \cite{Cardy}. The geometry of these D-branes was 
uncovered in \cite{AS,FFFS} where it was shown that Cardy's boundary states correspond
geometrically to D-branes wrapping conjugacy classes of the group manifold.
Additional D-brane states preserving an automorphism twisted symmetry
were constructed in \cite{BFS} and shown to wrap twisted conjugacy classes
\cite{FFFS}. The non-commutative gauge theory living on their world volume
was further investigated in \cite{AFQS}.
Permutation branes \cite{Reck,MGSSN}, where the automorphism is the exchange
of the  symmetry algebras of the two factors, are a special class of such
branes, see also the generalization discussed in \cite{FQ,FG}.

Repeating a similar program for unoriented strings, one can relate
algebraic and geometric constructions of orientifolds.
The crosscap states preserving the full symmetry algebra are those of
\cite{PSS}. Their geometric interpretation on group manifolds
has been investigated in \cite{Bru,HSS,BCW} with the result that
the orientifold planes are fixed point sets under an involution of the
group manifold and are localized on specific conjugacy classes. The basic
example is provided by the involution $g\to g^{-1}$, leading to orientifold
points sitting at elements of the center of the group whose conjugacy
class consists only of one point. 
This parity action can be modified by multiplication by an element $c$ of
the center $g \to cg^{-1}$. 
In CFT terms, this choice amounts to dressing the parity action 
by a simple
current \cite{HSSkb,FHSSW,BH1}. 

In the classical limit, the closed and open string states
have an interpretation in terms of functions on the group manifold
(closed strings)
or on the conjugacy classes wrapped by the D-brane (open strings). 
The parity acts as an involution
of the group target and induces an action on those function spaces. 
This geometrical action has been matched
with the one read off from conformal field theory one-loop
calculations in \cite{Bru,BCW}.

In this paper, we consider products of group manifolds and their
orientifold actions. The fixed point set of these orientifolds
are twisted conjugacy classes, just like in the case of D-branes.
We study the examples $SU(2)_k\times SU(2)_k$ and $U(1)_k\times U(1)_k$
in detail and 
check the localization of the orientifold plane by scattering  closed
string states off it \cite{MMS,HSS,Bru,BCW,GZ}. 
Furthermore, we consider the action of the geometric parity on
the space of functions living on the twisted conjugacy class and
compare its action with the one derived with the help of conformal field
theory. We find complete agreement in the classical (large $k$) limit of the
conformal field theory.

Having studied both $su(2)$ and $u(1)$ in detail, we move on to
the coset $su(2)_k\oplus u(1)_2/u(1)_{k+2}$. This is a GSO projected version
of the $N=2$ supersymmetric minimal model which has an alternative
description in terms of a Landau-Ginzburg model with
superpotential $W=X^{k+2}$.  For $N=2$ theories, there are two
types of orientifolds, namely A-type (those that for sigma-models correspond
to fixed point sets of an anti-holomorphic involution) and B-type 
(corresponding to holomorphic involutions). For a single minimal model,
the crosscaps and parities have been studied in \cite{BH2}, and
results from the Landau-Ginzburg and conformal field theory analysis,
such as the action of the parity on D-branes and open string states, have
been found to agree. One of the main applications for supersymmetric
minimal models is the construction of exact string vacua via Gepner models,
whose orientifolds have been studied in \cite{Gepner}. 

Concerning the extension to permutation orientifolds, 
the techniques discussed so far in this
paper are directly applicable to the (non spin-aligned) 
tensor product of two coset models, each with a separate GSO projection.
However, one would really be interested in the (GSO projected) tensor product
of two spin aligned $N=2$ minimal models, which is what we
consider in section \ref{SUSY}, focussing on the B-type case. 
We observe that the parity action
on open and closed string states is to some extend inherited from
that of the constituent $su(2)_k$ and $u(1)$ theories. However, there
are a number of choices of how the parity can act on the $u(1)_2$ part,
describing fermions and spin structures. In particular, for any
parity action one can consider a related parity that differs in its
action on the D-brane from the
initial one by a brane-anti-brane flip. In theories with a geometrical
interpretation, one would interpret this as an orientation reversal.

We also consider the Landau-Ginzburg description of the tensor
product, which has a superpotential $W=X_1^{k+2}+ X_2^{k+2}$.
In \cite{HWLG} it was explained how to construct
topological B-type crosscap states  making
use of the matrix factorization techniques of
\cite{KLII,HL}. We apply their techniques to the case
of the tensor product of two $N=2$ minimal models.
Following \cite{HWLG} one can
in particular derive from the Landau Ginzburg point of view
how the parity associated to a factorization 
acts on the B-type D-branes. The latter  are also
described by matrix factorizations, following the
ideas of \cite{MK}.  
We compare these results with the conformal field theory analysis
and identify the conformal field theory description of 
the parity that is most natural in the Landau Ginzburg
model.

We do not consider permutation orientifolds of Gepner models, which
are however covered in the paper \cite{hosomichi}.

\bigskip

\noindent
{\bf Note added:} Some of the results 
of this paper were independently obtained
in the paper \cite{hosomichi} by Hosomichi.

\section{Permutation boundary and crosscap states}

The tensor product of two identical rational conformal field theory
models carries a natural action of the permutation group. Accordingly,
one can use this symmetry to twist the gluing conditions for
boundary and crosscap states. 

We will consider the case of the charge conjugation modular invariant.
The symmetry algebra of the model is $\cA^1 \otimes \cA^2$, where 
$\cA^i =\cA^i_L \otimes \cA^i_R$, and $\cA^i_L=\cA^i_R=\cA$. The generators
of $\cA$ are denoted as $W$. 
The Hilbert space of the theory is then
\beq
\cH = \bigoplus_{i,j} \left( \cH_i \otimes \cH_{\bar{i}} \right)
\otimes \left( \cH_j \otimes \cH_{\bar{j}} \right) \ ,
\eeq
where $i,j$ are representations of $\cA$.
As usual, one can consider D-branes described by Cardy's boundary
states that preserve the full symmetry algebra. The corresponding
gluing conditions on the upper half plane are
\beq\label{untwisted}
W^{(1)}(z) = \oW^{(1)}(\bar{z}), \quad W^{(2)}(z) = \oW^{(2)}(\bar{z})
\quad {\rm for} \quad z=\bar{z} \ .
\eeq
Making use of the permutation symmetry, one can define the following
twisted 
gluing conditions \cite{Reck} 
\beq
W^{(1)}(z) = \oW^{(2)}(\bar{z}) \quad {\rm  for} \quad z=\bar{z}
\eeq
In the closed string channel, these gluing conditions are implemented
on coherent Ishibashi states which fulfill
\beq\label{permbc}
\left( W_n^{(1)} - (-1)^{s_W} \oW_{-n}^{(2)} \right)\kket{(i,i)}_{(12)}=0 \ ,
\eeq
where the subscript $(12)$ reminds us that this is the Ishibashi
state for the permutation boundary conditions.
If we let $i$ label the chiral primaries of a single factor, then the
permutation Ishibashi states can be built on states in the 
sector $\cH_i \otimes \cH_{\bar{i}} $, as in \cite{Reck}. Explicitly, they are written 
\begin{equation}
\kket{(i,i)}_{(12)}:=\sum_{M,N}|i,M\rangle\otimes|i,N\rangle\otimes 
|\bar{i},N\rangle\otimes |\bar{i},M\rangle
\end{equation}
More generally, we can have a pair of indices $(i_1,i_2)$ labeling the tensor product of representations
of the symmetry algebra for the left movers, while the summation over $M,N$ runs over the descendants.
Permutation Ishibashi states
can only be built on ground states whose labels $i_1,i_2$ agree.
The inner product between twisted Ishibashi states is given by
\begin{equation}\label{permproduct}
{}_{(12)}\bbra{(i,i)}e^{2\pi i \tau H_c}\kket{(j,j)}_{(12)}=
\delta_{ij}\chi_i(2\tau)^2
\end{equation}
and the one between twisted and untwisted Ishibashi states 
corresponding to  (\ref{untwisted}) by
\beq\label{twined}
{}_{\one}\bbra{(i_1,i_2)}e^{2\pi i \tau H_c}\kket{(j,j)}_{(12)}=
\delta_{i_1,j} \delta_{i_2,j} \chi_j(4\tau) \ .
\eeq
Permutation boundary states then take the form \cite{Reck}
\beq\label{permbs}
\left|\Scr{B},J\right\rangle_{(12)} = \sum_j \frac{S_{Jj}}{S_{0j}} \ \kket{(j,j)}_{(12)} \ .
\eeq
This is a special case of a more general construction described in \cite{GaGa},
taking a novel point of view on \cite{BFS}. Given an automorphism $\omega$
of the chiral algebra, one can consider the $\omega$-twisted boundary
conditions, (for the case that $\omega$ is a permutation this is 
(\ref{permbc})).
One can then construct $\omega$ twisted Ishibashi states and determine the
inner product between two twisted Ishibashi states, which results in
characters describing the propagation of strings in the tree-channel.
For the case of the permutation automorphism, this is just (\ref{permproduct}).
However, in addition there are strings propagating between 
two branes where the gluing conditions have been twisted by different
automorphisms. In particular, one can consider one of the automorphisms
to be trivial, leading to Cardy's boundary states, which preserve the
full symmetry. 
The overlap of an $\omega$ twisted Ishibashi state and an untwisted
Ishibashi state is given by a {\it twining character},
{\it i.e.} a trace over a Hilbert space with an insertion of the induced
action of $\omega$ on that Hilbert space
\beq
\chi_{\mu} (\tau) = \Tr_{\H_{\mu}} \big(\tau_\omega \ 
e^{2\pi i \tau(L_0-\frac{c}{24})} \big) \ .
\eeq
In our case, this is just (\ref{twined}).
Twisted boundary states can be written as linear combinations of
twisted Ishibashi states.
The coefficients, with which the twisted Ishibashi states are combined
to consistent boundary states are restricted by the Cardy condition.
To analyze this condition
between twisted and untwisted boundary states, one needs to transform
the tree level amplitudes to the open string channel, where in
particular one makes use of the modular transformation properties of the
twining characters \cite{Kac,FSS}.  It turns out
that one obtains integer combination of suitable twisted characters in the open
string channel if one chooses as coefficients the S-matrix elements for 
the twining characters, divided by the square-root of ordinary S-matrices
as usual. The integers appearing in the open string channel
are then the twisted fusion rule coefficients, which describe the
fusion of a twisted representation with an untwisted one, resulting
in a twisted representation, see \cite{GaGa} for
details.

In our case, the twining characters appearing in the closed string channel
are simply (\ref{twined}), and their modular transformation is
performed using the ordinary $S$-matrix of a single model. The coefficients
of the permutation boundary state (\ref{permbs}) are precisely that $S$-
matrix divided by the normalization $S_{0j}$, following the pattern
described above.

Turning to orientifolds, one would similarly like to consider parity actions
which involve the exchange of the two symmetry algebras.
In the closed string channel, the conditions on the crosscap states can
be obtained by conjugating (\ref{permbc}) with $e^{\pi i L_0}$, where
$L_0=L_0^{(1)}+L_0^{(2)}$.
This results in the following condition on
crosscap Ishibashi states
\beq
\left(W_n^{(1)} - (-1)^{s_W+n} \oW_{-n}^{(2)} \right)\cket{(i,i)}_{(12)}=0 \ .
\eeq
The crosscap Ishibashi states are obtained from the boundary
ones as
\beq
\cket{(i,i)}_{(12)}=
e^{\pi i (L_0-h^{tot}_i)}\kket{(i,i)}_{(12)} \ .
\nonumber
\eeq
The closed string amplitudes between permutation boundary and crosscap
states are
\begin{eqnarray}
{}_{(12)}\cbra{(i,i)} e^{2\pi i \tau H_c} \cket{(j,j)}_{(12)} &=&
\delta_{ij} \chi_i(2\tau)^2 \\ \nonumber
{}_{(12)}\bbra{(i,i)}e^{2\pi i \tau H_c}\cket{(j,j)}_{(12)}
&=& \delta_{ij}\hat\chi_j(2\tau)^2 \\ \nonumber
{}_{\one}\bbra{(i_1,i_2)}e^{2\pi i \tau H_c}\cket{(j,j)}_{(12)} &=&
\delta_{i_1,j} \delta_{i_2,j} \hat\chi_j (4\tau) \ ,
\end{eqnarray}
where $\hat\chi(\tau) = e^{-\pi i (h_i-\frac{c}{24})} \chi(\tau+\frac{1}{2})$
as usual.
The form of the permutation boundary state suggests the following ansatz
for the permutation crosscap state
\beq\label{pcross}
|\Scr{C}, \mu \rangle_{(12)} = \sum_j \frac{S_{\mu j}}{S_{0j}} \cket{(j,j)}_{(12)}=
\sum_j e^{2\pi i Q_{\mu}(j)} \ \cket{(j,j)}_{(12)}
\eeq
where $\mu$ is a simple current and $Q_{\mu}(j)$ the monodromy charge 
of $j$ with respect to $\mu$.

It is interesting to ask if there is an extension of the general construction
of twisted boundary states of \cite{BFS,GaGa} to the crosscap case.
The natural ansatz for a crosscap state would then be to combine the
Ishibashi states to full crosscap states using the modular matrix that
relates the closed string channel of the {\it mixed} amplitude between a
twisted crosscap state and an untwisted boundary state to the open string
channel. It is immediately clear, that the standard PSS crosscaps \cite{PSS}
fall into this pattern: They are given by
\beq \label{PSScross}
\vert \Scr{C}, \mu \rangle_\one = \sum_i \frac{P_{\mu i}}{\sqrt{S_{0i}}}
\cket{i} \ ,
\eeq
where as before $\mu$ is a simple current. The matrix $P=\sqrt{T}ST^2S\sqrt{T}$ appearing as
a prefactor of the Ishibashi state is precisely the one relating open
and closed string channel of the untwisted M\"obius strip.

What we can seen here is that also the permutation crosscap states
obey this construction prescription. Namely,
if we consider a mixed amplitude between a permutation
crosscap state and a Cardy boundary state, the hatted characters appearing
in the closed string channel are transformed to the open string channel
using the matrix $S$ for a single factor. The prefactors appearing
in the formula for the permutation
crosscap state are thus natural from this point of view.

We have listed all one-loop amplitudes involving the permutation crosscap
state in  appendix A. Our formulas show that the coefficients appearing in
the loop amplitudes are integers, thus providing necessary conditions for
the consistency of the crosscap state.

Let us highlight a few amplitudes that are of special interest.
There is a special permutation D-brane carrying the label $J=0$, where $0$
refers to the vacuum representation. The Cardy brane corresponding to this
label would have only a single vacuum character that appears in the open string
sector.
In the case of the permutation brane, the cylinder amplitude of the $J=0$
permutation brane with itself takes the form \cite{Reck}
\beq\label{cardypermutation}
{}_{(12)} \langle \Scr{B},0 \vert e^{-\frac{\pi i H_{c}}{\tau}} \vert \Scr{B},0 \rangle
_{(12)}=:\cC_{(12)(12)}(0,0)=\sum_j \chi_j (\tau) \chi_{\bar{j}}(\tau)
\eeq
and ``coincides'' with the diagonal bulk partition function 
if one identifies the right movers with the boundary fields in the second
tensor product factor \cite{Reck}. 
In the case where a sigma-model interpretation of
the theory is available, this brane should describe a brane whose worldvolume
is the conjugate diagonal in $M\times M$. 
We now calculate the M\"obius amplitude of this D-brane with the 
permutation crosscap state
\beq\label{cardypermutationmoebius}
{}_{(12)} \langle \Scr{C},0 \vert e^{-\frac{\pi i H_{c}}{4\tau}} \vert \Scr{B},0 \rangle
_{(12)}=:\cM_{(12)(12)}(0,0) = \sum_j \hat\chi_j(\tau) \hat\chi_{\bar{j}}(\tau),
\eeq
which shows that all open string states are invariant under the orientifold
action. We can therefore conclude that our crosscap state corresponds to
an orientifold plane that is located on the diagonal. Turning the
argument around, we could have postulated that the orientifold 
we are looking for leaves the brane on the diagonal invariant and acts trivially on all
its open string states. By a modular transformation, we could then have
concluded that (\ref{pcross}) is the corresponding crosscap state.

\medskip
\noindent
We will use the notation introduced in (\ref{cardypermutation}) and 
(\ref{cardypermutationmoebius}) throughout
the paper: $\cC$, $\cM$ and ${\cal K}$ stand for the loop channel of
the cylinder, M\"obius and Klein bottle amplitude respectively.
The subscripts refer to the automorphism type of the boundary and
crosscap states, whereas the two arguments given in brackets are
the Cardy labels of the boundary or crosscap states.

\section{Group manifolds}
A special example are the cases of group manifolds $H$ \cite{FFFS,
AFQS,FS}. Here, boundary states corresponding to boundary conditions 
twisted by an automorphism $\omega$ have been constructed in \cite{BFS} and
interpreted in terms of branes wrapping twisted conjugacy classes 
$\cC_\omega$ in \cite{FFFS}
\beq
\cC_{\omega}(g) = \{ (h^{-1} g \omega (h) \vert h \in H \} \ .
\eeq
For the special case of products of group manifolds $H=G\times G$ where
$\omega$ acts as the exchange of the two factors, one obtains
\beq
\cC_{(12)}((g_1,g_2)) = \{ (h_1^{-1} g_1 h_2, h_2^{-1} g_2 h_1) \ \vert \
h_i \in G \} \ .
\eeq
The brane corresponding to the boundary state $J=0$ is the twisted
conjugacy class of the identity 
\beq\label{twistedone}
\cC_{(12)}((\one,\one)) = \{(h,h^{-1}) \vert h \in G \}
\eeq
The orientifold actions we want to consider are supposed to exchange
the left moving current of the first WZW model with the right moving
one of the second. The currents are given by
\beq
J^{(i)}(z) = k g_i^{-1} \partial g_i \quad \quad \bar{J}^{(i)}(\bar{z}) = 
-k \bar\partial g_i g_i^{-1} , \quad \quad i=1,2 ,
\eeq
and the possible orientifold actions involve world sheet parity $\Omega$
combined with an action on the group manifold $g^{(1)} \to (g^{(2)})^{-1}$. 
This action
can be modified with a translation by an element of the center of the
group, corresponding in conformal field theory to the dressing by
the action of a simple current, reflected in the label $\mu$ in (\ref{pcross}).

The basic orientifold action will leave the twisted conjugacy class of the 
identity 
(\ref{twistedone}) pointwise invariant, and the orientifold fixed point plane
is located in the same place.
This is in complete agreement with the discussion of the one-loop amplitude.
Similar statements hold for the conjugacy classes $(c,c)$ where $c$ is an
element of the center. 

The geometry of the other 
conjugacy classes $\cC_{(12)}((g_1,g_2))$ with $(g_1,g_2) \neq (\one, \one)$
or $(c,c)$
has been discussed in \cite{FS}, which we will now review.
There is a natural surjection $m: G\times G \to G$ given by group multiplication, $m(g_1,g_2)= g_1g_2$. It follows immediately that the twisted conjugacy classes wrapped  by the permutation branes get mapped
to ordinary conjugacy classes of $G$. Indeed, one can also easily see the
opposite statement, namely that if $g_1$ and $g_2$ are conjugate in $G$
then their preimages $(g_1',g_1'')$ with $g_1'g_1''=g$ and
$(g_2',g_2'') $ with $g_2' g_2''=g_2$ in $G\times G$ are 
in the same twisted conjugacy class in $G\times G$. The conclusion \cite{FS}
is that the twisted conjugacy classes are precisely the inverse
images under $m$ of the conjugacy classes in $G$. They are principal 
$G$-bundles over the conjugacy classes of $G$.

In particular, one
sees from this point of view that the corresponding boundary states carry
the same set of labels as the ordinary Cardy boundary states describing
branes wrapping conjugacy classes in $G$. This is of course in complete
agreement with labelling of the permutation boundary states from the
conformal field theory point of view (\ref{permbs}). 
In addition, we can  conclude that the orientifold 
$g_1\to g_2^{-1}$ will leave a
twisted conjugacy class $(g_1,g_2)$ invariant if $ g=g_1 g_2$ is
conjugate to its inverse.  However, in general the conjugacy
class will not be pointwise fixed. As opposed to the case of
Cardy branes on a single factor, orientifold actions dressed by an
application of an order $2$ element of the center will map the
twisted conjugacy classes (set-wise) in the same way, though might act differently
on the individual elements of the class. We will see examples for
this in the case of $SU(2)\times SU(2)$ which we will now discuss in
detail.

\subsection{$SU(2)\times SU(2)$}

Let us consider the case $SU(2) \times SU(2)$. In this case, we will
be able to give an interpretation for the orientifold action for all
D-branes. 

First, we will summarize the results of \cite{BCW,HSS,Bru}
on the orientifolds of a single $SU(2)$ model. In this case, there are two
orientifolds to consider, parity inversion with a combination of 
$g\to \pm g^{-1}$. In conformal field theory, this corresponds to the
fact that there are two order two simple currents in the model, the
identity and the current corresponding to the representation $k/2$.
The crosscap states are then given as in (\ref{PSScross}) \cite{PSS},
where $\mu=0$ or $\mu=k/2$.

We start by discussing the involution $g\to +g^{-1}$, corresponding to
the choice $\mu=0$ on the CFT side.
The fixed point set consists of the two group elements $\pm \one$ such
that this parity describes a $0$-dimensional orientifold. The two orientifold
points can have equal or opposite tensions, and, using the properties of the
$SU(2)$ $P$-matrix, it is easy to see that rational conformal field theory
realizes the case of opposite tensions (leading to vanishing total tension)
for the case $k$ odd, and equal tension for the case $k$ even. A priori,
this tension could be positive or negative, corresponding to an overall
sign by which the crosscap can be multiplied.

Turning to the open string sector, D-branes on $SU(2)$ wrap conjugacy
classes. Two conjugacy classes, namely those of $\pm \one$ are pointlike,
the others are $2$-spheres. The map $g\to g^{-1}$ induces an inversion
on each individual two-sphere. 

The ground states of the open string
spectrum are spherical harmonics, whose behavior under inversion is standard
\beq
Y_{lm} \to (-1)^{l} Y_{lm} \ .
\eeq
This determines the projection in the open string channel up to an overall sign
that can depend on the specific brane and the overall sign of the crosscap.

These geometrical expectations are nicely matched by the explicit evaluation
of the conformal field theory M\"obius strips, with the result
\beq
\langle \Scr{C},0 | e^{-\frac{\pi i}{4\tau} H_{c}} | \Scr{B}, J \rangle
= (-1)^{2J} \sum_{j=0}^{{\rm min} \{ 2J, k-2J \} } (-1)^j \chi_j(\tau) \ ,
\eeq
where $J$ labels the Cardy boundary state and the sum is taken
over integer $j$. The
special values $J=0$ and $J=k/2$ correspond to the pointlike conjugacy classes,
all other values label two-spheres \cite{AS,FFFS}. 

The other involution $g\to - g^{-1}$ exchanges the two
pointlike conjugacy classes and leaves the equatorial $2$ sphere fixed. 
Hence, one expects the involution to act trivially on the open string
states on the brane wrapping the equatorial two-sphere. This can be
verified from the conformal field theory point of view for the case $k$ even,
where the brane with $J=k/4$ wraps the equator. All other branes
are mapped to image branes by a reflection at the equator, in terms
of conformal field theory, the image of the brane $J$ is $k/2-J$ as
is reflected in the M\"obius amplitude
\beq
\langle \Scr{C},\frac{k}{2} | e^{-\frac{\pi i}{4\tau} H_{c}} | 
\Scr{B}, J \rangle = 
\langle \Scr{C},\frac{k}{2} | e^{-\frac{\pi i}{4\tau} H_{c}} | 
\Scr{B}, \frac{k}{2}-J \rangle = \sum_{j=\frac{k}{2}-2J}^{k/2} \chi_j(\tau) \ .
\eeq
To generalize this to the case of twisted conjugacy classes of a 
product of two $SU(2)$'s, 
we have to understand the algebra of functions on
the twisted conjugacy class, following \cite{AFQS}. It is then
possible to interpret the orientifold action derived from the M\"obius
strip in terms of an involution of that algebra, giving a geometric
interpretation of the orientifold action. 
The authors of \cite{AFQS} consider the open string sector for an 
arbitrary $\omega$-twisted
D-brane on a compact simply connected simple group manifold $H$, where $\omega$
denotes the automorphism. We want to apply their strategy to the case
$H=G\times G$, and $\omega$ the permutation of the two factors. 
The
twisted boundary conditions of \cite{AFQS}
are labelled by representations of the $\omega$-
invariant subgroup $H^\omega=\{ h\in H \vert \omega(h)=h \}$. In our case
$H^{\omega}$ is the diagonal subgroup $G$ of $G\times G$ and we have seen
that indeed the permutation D-branes carry representation labels of $G$.
Quite generally, the open string sector for a pair of 
twisted conjugacy classes labelled $(J_1,J_2)$ is realized  
by the $H$-module ${\cal A}^{(J_1,J_2)}$, where
\beq
{\cal A}^{(J_1,J_2)} \sim {\rm Inv}_{H^\omega} 
\left( {\cal F}^{(J_1,J_2)}\right) \quad {\rm and} \quad
{\cal F}^{(J_1,J_2)}:= {\cal F}(H) \otimes {\rm Hom} (V_{J_1}, V_{J_2}) \ .
\eeq
Here, ${\cal F}(H)$ denotes the algebra of functions on the group $H$ and
$V_{J_i}$ is a representation space for the irreducible representation $J_i$.
The group $H\times H$ acts on the space of functions 
${\cal F}(H)$ by the regular action so that we have a natural action
of $H \times H^{\omega}$ on the space of matrix valued functions 
${\cal F}^{(J_1,J_2)}$ by
\beq
A^{h_1,h_2}(g) = R_{J_2}(h_2) A(h_1^{-1} g h_2) R_{J_1}(h_2)^{-1},
\eeq
where $h_1 \in H$ ($=G\times G=SU(2) \times SU(2)$ in our case) and 
$h_2 \in H^\omega$ (= $G$ 
in our case) and $R_{J_i}$ are representation matrices of $h_2$. 
The space ${\cal A}^{(J_1,J_2)}$ is then the restriction to those matrix
valued functions that are invariant under the action of 
${\one} \times H^\omega$ which is ${\one} \times G$ in our case. 
To show that this $H$-module is equivalent to the module of open string
ground states determined by the open string partition function, \cite{AFQS}
decompose ${\cal A}^{(J_1,J_2)}$ into irreducibles.
In our case, using the Peter-Weyl theorem
\beq
{\cal F}(G\times G) = {\cal F}(G) \otimes {\cal F}(G)
\equiv \left( \bigoplus_{j_1} V_{j_1} \otimes V_{j_1} \right) \otimes
\left( \bigoplus_{j_2} V_{j_2} \otimes V_{j_2} \right),
\eeq
where $j_i$ label irreducible representations of $SU(2)$. 
This space, which is a representation space of $(G\times G)^2$
has to be decomposed with respect to the diagonal $G$ acting in the right
regular action, which is easily achieved using the non truncated fusion rules of the Lie group.
To obtain ${\cal F}^{(J_1,J_2)}$ the result has to be tensored
with $V_{J_1} \otimes V_{J_2}$,
\beq
{\cal F}^{(J_1,J_2)}= \bigoplus_{j_1, j_2, j} N_{j_1 j_2}^j V_j \otimes V_{j_1}
\otimes V_{j_2} \otimes V_{J_1} \otimes V_{J_2} \ .
\eeq
To find ${\cal A}^{(J_1,J_2)}$ one has to reduce to the $G$-invariant part,
which can again be done using the fusion rules. The result is
\beq
{\cal A}^{(J_1,J_2)}= \bigoplus_{j_1, j_2, j} N_{j_1 j_2}^j N_{J_1J_2}^{j} V_{j_1}
\otimes V_{j_2}.
\eeq
Comparing to the open string partition function between
permutation branes labelled $J_1$ and $J_2$ 
\beq
\cC_{(12)(12)}(J_1,J_2) = \sum_{j,j_1,j_2} N_{J_1J_2}^j N_{j_1 j_2}^j
\chi_{j_1}(\tau) \chi_{j_2} (\tau)
\eeq
we see that these considerations
have correctly reproduced the  structure of open string ground states.
As we
have seen \cite{FS}, 
the twisted conjugacy classes are principal G-bundles over the
conjugacy classes of $G$. Hence, they locally look like a conjugacy class
times the group itself, and for the $SU(2)$ case this is true also globally,
so that the geometry of the twisted conjugacy class is $S^2 \times S^3$. 
One would expect that the parity acts on $S^3$ either as the identity
or the $\ZZ_2$ anti-podal identification. 
For the branes wrapping the conjugacy classes $S^3\times \{ pt \}$ one
therefore expects a trivial action in the first case, which motivates that
the M\"obius strip should take the form
\beq
\cM_{(12)(12)}(0,0)= \sum_{j=0}^{\frac{k}{2}} \hat\chi_j(\tau) \hat\chi_j(\tau) \ .
\eeq
This, as mentioned before, is indeed the M\"obius strip for the crosscap
corresponding to the trivial simple current  $\mu=0$.
In the second case one would expect that the open string ground states
transform in the same way as when taking a $\ZZ_2$ orbifold of a single $SU(2)$
to obtain $SO(3)$, that is the expectation for the M\"obius strip is
\beq
\cM_{(12)(12)}(\frac{k}{2},0)= \sum_{j=0}^{\frac{k}{2}} (-1)^{2j} \hat\chi_j(\tau) \hat\chi_j(\tau) ,
\eeq
which is indeed the M\"obius strip involving the crosscap 
with $\mu=\frac{k}{2}$. 
For the other D-branes, 
wrapping  conjugacy class of the topology $S^3\times S^2$, this action
has to be combined with the action coming from the $S^2$ part. It can be
expected that the latter is simply inherited from the action of the
inversion on the conjugacy classes of a single $SU(2)$ factor.
As reviewed
at the beginning of this section, the latter
is an inversion under which the spherical harmonics pick up a sign $(-1)^j$,
where $j$ is integer.
This leads to the following geometrically motivated
expectation for the M\"obius strip involving the brane labelled $J$
for the case $\mu =0$
\beq
\cM_{(12)(12)}(0,J)= \sum_{j,j_1,j_2=0}^{\frac{k}{2}} N_{j_1 j_2}^j N_{JJ}^j (-1)^{2J+j} \hat\chi_{j_1}(\tau)\hat\chi_{j_2}(\tau),
\eeq
where we have included the same brane-dependent sign $(-1)^{2J}$ that
appeared in the M\"obius amplitudes for a single $SU(2)$.
On the other hand, the M\"obius strip derived from conformal field theory
is
\beq
\cM_{(12)(12)}(0,J)= \sum_{j_1, j_2=0}^{\frac{k}{2}} Y_{J j_1}^{j_2} \hat\chi_{j_1}(\tau) \hat\chi_{j_2}(\tau),
\eeq
where we refer to (\ref{Y definition}) in the appendix for the definition of the $Y$ tensor.
In order to obtain an agreement between these two expressions, we need the identity, (for $J,j_1,j_2\leq k/4$)
\beq\label{YNN}
Y_{J j_1}^{j_2} = \sum_{j=0}^{\frac{k}{2}} N_{JJ}^j N_{j_1 j_2}^j (-1)^{2J+j},
\eeq
which can indeed be shown to be true using the explicit form for the $SU(2)$
$Y$-tensor given in the
appendix.
Likewise, the geometrical expectation for the amplitude between the brane $J$
and the crosscap with label $\mu=\frac{k}{2}$ is
\beq
\cM_{(12)(12)}(\frac{k}{2},J)= \sum_{j,j_1,j_2=0}^{\frac{k}{2}} N_{j_1 j_2}^j N_{JJ}^j (-1)^{2j_1+j+2J} 
\hat\chi_{j_1}(\tau)\hat\chi_{j_2}(\tau) \ .
\eeq
The prefactor of $\hat\chi_{j_1}(\tau)\hat\chi_{j_2}(\tau)$ derived from conformal field
theory is $Y_{\frac{k}{2}-J,j_1}^{j_2}$, and to get agreement 
(up to a factor of $(-1)^k$) 
we use (\ref{YNN}) and the following symmetry of the
$Y$-tensor of $SU(2)$ (see appendix B) 
\beq
Y_{\frac{k}{2}-J,j_1}^{j_2}= (-1)^{k+2j_1}Y_{Jj_1}^{j_2} \ .
\eeq
One might wonder if there also exists a crosscap state exchanging the
permutation boundary states $J=0$ and $J=\frac{k}{2}$, similar to what happened
in the case of a single $SU(2)$. This would imply that the conjugacy class
of $(\one,\one)$ gets mapped to $(\one, -\one)$, as would happen under
the map $g_1 \to - g_2^{-1}$, $ g_2\to g_1^{-1}$. This parity squares to
$-\one$ and in particular is non-involutive. That means that the
corresponding crosscap states have to be built on circles twisted by 
$P^2\neq \one$.
In our case  the ground states on which the twisted permutation crosscap state
are built differ in the first and second factor by an application of
the simple current $\frac{k}{2}$. The crosscap Ishibashi states
take the form
\beq
\cket{(j,\frac{k}{2}-j)}_{(12),P^2}=e^{\pi i(L_{o}^{tot}-h_{j}^{tot})}\sum_{M,N=0}^{\infty}\vert j,M \rangle\otimes\vert \frac{k}{2}-j,N \rangle \otimes U\vert \frac{k}{2}-j,N \rangle \otimes U\vert j,M \rangle \ .
\eeq
This leads to the following inner product between
such permutation crosscap states
\beq
{}_{(12),P^2}\cbra{(j,\frac{k}{2}-j)}e^{-\frac{\pi i H_c}{2\tau} }
\cket{(j',\frac{k}{2}-j')}_{(12),P^2} = \delta_{j,j'} \chi_j(-\frac{1}{2\tau})
\chi_{\frac{k}{2}-j}(-\frac{1}{2\tau}),
\eeq
where the subscript $P^2$ is supposed to indicate that the circle is
twisted by $P^2$.
The full twisted crosscap state
\beq
\vert \Scr{C} \rangle_{(12)P^2} = \sum_{j=0,j\in\mathbb{N}}^{\frac{k}{2}} \Big( \cket{(j,j)}_{(12)}+\cket{(j,\frac{k}{2}-j)}_{(12)P^2}\Big)
\eeq
has a natural interpretation as a crosscap state in the simple current
orbifold leading to the $D$-modular invariant. For $k=0$ mod $4$ this is
a simple current extension. We have checked that this crosscap state
leads to consistent one-loop amplitudes, the
results can be found in \cite{mitev}. For a general description
of permutation crosscap states on orbifolds see \cite{hosomichi}.\\

To complete the geometrical picture of the permutation orientifold,
we show how the localization of the orientifold fixed planes can
be determined by scattering localized closed string wave packets,
following \cite {MMS}, see also \cite{FFFS,Bru,HSS,GZ}. 
For the case of permutation D-branes,
a similar calculation was done in \cite{GZ}. 

The permutation crosscap state associated to the identity simple current is given by the expression involving the Ishibashi crosscap states
\begin{eqnarray}
\left|\Scr{C},0\right\rangle_{(1 2)}&=&\sum_{j=0}^{\frac{k}{2}}
\frac{S_{0j}} {S_{0j}}\cket{j}_{(1\ 2)}\nonumber\\
&=&\sum_{j=0}^{\frac{k}{2}}e^{\pi i (L_0-h_j)}\sum_{M_1,M_2=0}^{\infty}\underbrace{\left|j,M_1\right\rangle\otimes  \left|j,M_2\right\rangle}_{\mbox{left}}\otimes\underbrace{ U\left|j,M_2\right\rangle\otimes U\left|j,M_1\right\rangle}_{\mbox{right}}\nonumber
\end{eqnarray}
We now take a closed string state $\left|g_1,g_2\right\rangle$ that is localized at $(g_1,g_2)\in $ SU(2)$\times$ SU(2). Its expansion coefficients are 
following \cite{MMS}, appendix D, given by
\begin{equation}
\label{localization equation}
\left \langle g_1,g_2\right|\big(\underbrace{\left|j_1,m_1,m_2\right\rangle}_{\mbox{ $1^{st}$ SU(2)}}\otimes \underbrace{ \left|j_2,n_1,n_2\right\rangle}_{\mbox{$2^{nd}$ SU(2)}}\big)\approx\sqrt{2j_1+1}\sqrt{2j_2+1}\mathcal{D}^{j_1}_{m_1,m_2}(g_1)\mathcal{D}^{j_2}_{n_1,n_2}(g_2)\nonumber
\end{equation}
where the $\mathcal{D}^j_{m,n}(g)$ are the matrix coefficients of the group element $g$ in the $2j+1$ dimensional representation of SU(2). Here the indices $m_1,m_2$ label only descendants of the horizontal subalgebra, meaning that we only act with zero modes.\\
If we now look at the coupling between the crosscap and this localized state and take the approximation of dropping  all the descendants appearing in the expression for $\left|\Scr{C},0\right\rangle_{(1\ 2)}$, then we get
\begin{eqnarray}
\left \langle g_1,g_2\right.\left|\Scr{C},0\right\rangle_{(1\ 2)} &\approx& \sum_{j=0}^{\frac{k}{2}}\sum_{M_1,M_2}\left \langle g_1,g_2\right|(\left|j,M_1,M_2\right\rangle\otimes \left|j,M_2,M_1\right\rangle)\nonumber\\
&=& \sum_{j=0}^{\frac{k}{2}}\sum_{M_1,M_2}(2j+1)\mathcal{D}^{j}_{M_1,M_2}(g_1)\mathcal{D}^{j}_{M_2,M_1}(g_2)\nonumber\\&=&\sum_{j=0}^{\frac{k}{2}}(2j+1)\mbox{Tr}(\mathcal{D}^{j}(g_1)\mathcal{D}^{j}(g_2))=\sum_{j=0}^{\frac{k}{2}}(2j+1)\mbox{Tr}(\mathcal{D}^{j}(g_3)) \, \nonumber
\end{eqnarray}
where $g_3=g_1g_2$ by the representation property of the $\mathcal{D}^j$ matrices. If we parametrize SU(2) by the polar angles $(\psi,\theta,\phi)$, then we obtain
\begin{equation}
\left \langle g_1,g_2\right.\left|\Scr{C},0\right\rangle_{(1\ 2)}\approx\sum_{j=0}^{\frac{k}{2}}(2j+1)\frac{\sin((2j+1)\psi_3)}{\sin(\psi_3)}=\sum_{n=1,n\in \mathbb{N}}^{k+1}n\frac{\sin(n\psi_3)}{\sin(\psi_3)}
\end{equation}
which as a function of $\psi_3$ has a strong peak at zero of height $\sum_{n=1}^{k+1}n^2=\frac{(k+1)(k+2)(2k+3)}{6}$ and is practically zero elsewhere. Thus we are interested at the points $(g_1,g_2)$ for which the latitude angle of their product is zero. Since these correspond to the north pole, we get the relation $g_1g_2=1$, so that the crosscap state is localized on the twined conjugacy class $\mathcal{O}_3^+=\left\{(g,g^{-1})|g\in \mbox{SU(2)}\right\}$ as was to be expected from our earlier considerations. This confirms that the permutation crosscap state $\left|\Scr{C},0\right\rangle_{(1\ 2)}$ associated to the identity simple current is localized on the fixed point set of the involution $(g_1,g_2)\mapsto (g_2^{-1},g_1^{-1})$.\\
Similarly we can look at the permutation crosscap associated to the simple current $\frac{k}{2}$. Since $S_{\frac{k}{2}j}=(-1)^{2j}S_{0j}$, this differs from the previous one only by a factor of $(-1)^{2j}$ in the expansion, meaning that the calculations are practically identical. One obtains
\begin{equation}
\langle g_1,g_2|\Scr{C},\frac{k}{2}\rangle_{(1\ 2)}\approx\sum_{n=1,n\in \mathbb{N}}^{k+1}(-1)^{n-1}n\frac{\sin(n\psi_3)}{\sin(\psi_3)}
\end{equation}
which no longer has a peak at zero but at $\psi_3=\pi$. Therefore one now has that $g_1g_2=-1$, meaning that the crosscap state is localized on the orientifold fixed plane $\mathcal{O}_3^-=\left\{(g,-g^{-1})|g\in \mbox{SU(2)}\right\}$ of the involution $(g_1,g_2)\mapsto (-g_2^{-1},-g_1^{-1})$.\\
In the language of twisted conjugacy classes that we developed previously, we have
\begin{equation}
\mathcal{O}_3^+=\cC_{(12)}((\one,\one))\quad \mbox{and}\quad \mathcal{O}_3^-=\cC_{(12)}((\one,-\one)) \ .
\end{equation}

\subsection{Rational $U(1)\times U(1)$ theories}
In this section, we consider D-branes on the torus $T=S^1 \times S^1$, where
both circles have equal radius $R=\sqrt{k}$, for a positive integer $k$. 
As is well known, in this situation one can extend the chiral algebra
by exponentials with integer weight, such that the theory becomes rational.
The Hilbert space of the theory is for the diagonal modular invariant
\begin{equation}
\cH = \bigoplus_{n_1,n_2 \in \ZZ_{2k}} 
\left( \cH_{n_1}\otimes \cH_{n_1} \right)
\otimes \left( \cH_{n_2}\otimes \cH_{n_2} \right) \ .
\end{equation}
There are two kinds of permutation orientifolds to consider, in the first
case the $U(1)$ current $J$ gets mapped to
\beq
P_B: J^{(1)} \to \bar{J}^{(2)}, \quad J^{(2)} \to \bar{J}^{(1)}
\eeq
and the corresponding permutation boundary conditions read 
\beq
J^{(1)}(z) = \bar{J}^{(2)}(\bar{z}), \quad J^{(2)}(z) = 
\bar{J}^{(1)}(\bar{z})
\eeq
The second case differs by a sign
\beq
P_A: J^{(1)} \to -\overline{J}^{(2)}, \quad J^{(2)} \to -\overline{J}^{(1)}
\eeq
and likewise for the boundary conditions.
We will in this paper refer to 
the two different choices of the parity or boundary conditions 
as A-type and B-type, in analogy to the $N=2$ supersymmetric
case, where a $U(1)$ current appears as part of the superconformal algebra.

Since $J^{(i)}= \sqrt{k} \partial X^{(i)}$, where $X^{(i)}$ is the free boson
parametrizing the $i$th circle, one can easily deduce that the branes 
and orientifold fixed planes are closed curves given by the equation
$X^{(1)} = \pm X^{(2)} + const$.
The permutation boundary state takes the form
\beq
|\Scr{B}, N\rangle_{(12)} = \sum_n \frac{S_{Nn}}{S_{0n}}\kket{(n,\pm n)}_{(12)}
=\sum_n e^{-\pi i \frac{Nn}{k}} \kket{(n,\pm n)}_{(12)}
\eeq
The sign $\pm$ depends on whether the state is A-type or B-type, apart from
that the discussion for the two cases is completely parallel. We already
know that geometrically these branes are wrapped on circles intersecting
the torus baseline at $45$ degrees, and the natural assumption is that
the label $N$ determines the intersection point of the D-brane with the
torus axes. \\
If we parametrize the torus by two angles $\phi_1,\phi_2\in[-\pi,\pi]$, we can apply the same localization procedure as before, just modified appropriately. We obtain 
\begin{equation}
\langle e^{i\phi_1},e^{i\phi_2}|\big(|m\rangle\otimes |-m\rangle\big)\approx e^{im(\phi_1-\phi_2)}
\end{equation}
so that the permutation boundary state with label $N$ is localized on the set 
\begin{eqnarray}
\langle e^{i\phi_1},e^{i\phi_2}|\Scr{B},N\rangle_{(1\ 2)}^B\approx\sum_{m=-k+1}^ke^{im(\phi_1-\phi_2-\frac{N \pi}{k})}\stackrel{k\rightarrow \infty}{\rightarrow}\mbox{const}\cdot\delta(\phi_1-\phi_2-\phi_{N})
\end{eqnarray}
where $\phi_{N}\in[-\pi,\pi]$ and $N$ has been taken to vary appropriately so that in the limit $k\rightarrow \infty$, $\frac{N \pi }{k}\rightarrow \phi_{N}$. Hence the permutation D-branes for the $B$-type boundary conditions are simply closed circles given by the equation $\phi_2=\phi_1-\phi_{N}$. The calculations for the $A$ type are analogous, with the result that $\phi_2=-\phi_1+\phi_{N}$. 
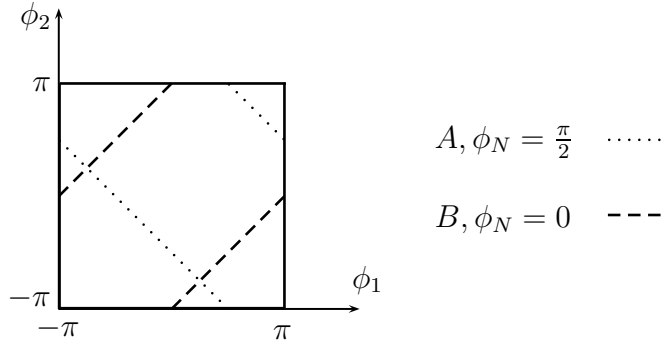
\begin{figure}[ht]
\begin{center}
 \begin{pspicture}(7,4) 
	 \put(3.9,0.2) {\makebox(0,0)[bl]{$\phi_1$}}
	 \put(-0.1,3.7) {\makebox(0,0)[br]{$\phi_2$}}
	 \put(-0.1,2.9) {\makebox(0,0)[br]{$\pi$}}
	 \put(3.1,-0.4) {\makebox(0,0)[br]{$\pi$}}
	 \put(-0.1,0) {\makebox(0,0)[br]{$-\pi$}}
	 \put(-0.3,-0.4) {\makebox(0,0)[bl]{$-\pi$}}
	 \psline[linewidth=0.75pt]{->}(0,0)(0,4)
	 \psline[linewidth=0.75pt]{->}(0,0)(4,0)
	 \psline[linewidth=1pt]{-}(0,0.01)(3,0.01)
	 \psline[linewidth=1pt]{-}(0.01,-0.01)(0.01,3)
	 \psline[linewidth=1pt]{-}(3,0.01)(3,3.01)
	 \psline[linewidth=1pt]{-}(0.01,3)(3.02,3)
	 \psline[linewidth=1pt,linestyle=dashed]{-}(0,1.5)(1.5,3)
	 \psline[linewidth=1pt,linestyle=dashed]{-}(1.5,0)(3,1.5)
	 \psline[linewidth=1pt,linestyle=dotted]{-}(0,2.25)(2.25,0)
	 \psline[linewidth=1pt,linestyle=dotted]{-}(2.25,3)(3,2.25)
	\put(5,1) {\makebox(0,0)[bl]{$B,\phi_{N}=0$}} \psline[linewidth=1pt,linestyle=dashed]{-}(7.3,1.25)(8,1.25)
	\put(5,2) {\makebox(0,0)[bl]{$A,\phi_{N}=\frac{\pi}{2}$}}  \psline[linewidth=1pt,linestyle=dotted]{-}(7.3,2.25)(8,2.25)
	\end{pspicture}
	\end{center}
	\caption{The loci of the A/B D-branes for different values i=of $\phi_N$}
	\end{figure}
As we shall see later on, the permutation crosscaps are localized on the same sets as their D-branes counterparts. 
The one-loop amplitude between two D-branes of equal type is
\beq
\cC_{(12)(12)} (N,M) = \sum_{n=-k+1}^k \chi_n(\tau) \chi_{N-M\mp n}(\tau),
\eeq
such that open string states for $N\neq M$ have non-trivial windings, which
is natural, since those branes are separated by a finite distance.
Likewise, we have the crosscap state
\beq
|\Scr{C},\mu \rangle_{(12)} 
=\sum_n e^{-\pi i \frac{\mu n}{k}} \cket{(n,\pm n)}_{(12)}
\eeq
Since the closed string couplings to the ground states are the same as
for the D-brane carrying the same label, an immediate conclusion that could
be reached by a calculation similar to the previous one is
that this crosscap state is located in the same place as the brane. Moreover, the one-dimensional orientifold fixed sets can be written as twisted conjugacy classes:
\begin{equation}
\mathcal{O}^A_1=\mathcal{C}_{(12)}((e^{i\phi_N},\mathbf{1}))\quad\mbox{and}\quad \mathcal{O}^B_1=\mathcal{C}_{(12)\circ I}((e^{i\phi_N},\mathbf{1})),
\end{equation}
where $I:e^{i\phi}\rightarrow e^{-i\phi}$ is the inversion.
The M\"obius amplitude between a crosscap and boundary state
of equal type is
\beq
\cM_{(12)(12)} (\mu,M) = \sum_{n} \hat\chi_{n} (\tau) 
\hat\chi_{2\mu-2M \mp n} (\tau) \sigma(\mu,M,n) 
\eeq
Here, $\sigma(\mu,M,n)$ is a sign taking
the values
\beq\label{sigmadef}
\sigma(\mu,M,n) = \left\{\begin{array}{ll}
1&\mbox{if $\mathcal{P}(2\mu-2M \mp n)=2\mu -2M \mp n$ mod $4k$}\\
(-1)^{n+k}&
\mbox{if $\mathcal{P}(2\mu -2M \mp n)=2\mu -2M \mp n+2k$ mod $4k$}.
\end{array}
\right.
\eeq
where $\mathcal{P}(n)=\hat{n}$ sends the argument to the standard range $\{-k+1,\ldots,k\}$. On general grounds, the multiplicities of the characters
appearing in the loop channel between permutation crosscap and boundary
state are given by the $Y$ tensor, which we have spelled out here
explicitly.

By comparing the Cylinder and M\"obius strip we see that the image brane
of the brane $M$ is the brane $2\mu-M$. In particular, a brane that remains
fixed under the involution has to fulfill $M=2\mu -M$ mod $2k$ with the two
solutions $M=\mu$ and $ M=\mu+k$. Note that in the two cases in 
(\ref{sigmadef}) are exchanged under $M \to M+k$. In particular, there
are no signs for the case $M=\mu$, but a non-trivial orientifold action 
on the open string states in the case
$M=\mu +k$. This indicates that the brane $M=\mu$ is left pointwise
fixed, as we discussed earlier. On the other hand,  
the brane $M=\mu+k$ is left only setwise fixed, the orientifold involution
identifies diametrically opposite points of the fixed point circle. The
open string states transform accordingly.

\section{The $N=2$ minimal models} \label{SUSY}

\subsection{The conformal field theory approach} \label{SUSYCFT}

We consider the tensor product of two $N=2$ supersymmetric minimal models.
A single minimal model with a non-chiral GSO projection has a description in
terms of a coset model $su(2)_k \oplus u(1)_2 / u(1)_{k+2}$. 
The representations of the coset algebra are labelled by $(j,n,s)$, where
$j$ denotes the spin of the $su(2)$ representation and 
$n\in \ZZ_{2k+4}$, $s\in \ZZ_4$. These labels
are subject to a selection rule that requires $2j+n+s$ to be even. Furthermore, one has
to identify
\beq
(j,n,s) \sim (\frac{k}{2}-j, n+k+2, s+2) \ .
\eeq
We will denote the equivalence classes by $[j,n,s]$. 
We choose the range  for the labels of the $U(1)_{k+2}$ theory to
be $-k-1, \dots, k+2$ 
and  denote 
the sum of $n_1$ and $n_2$ reflected back into that  range
 by $n_1\hat{+} n_2$.
Likewise, our  range for the $s$ labels is $-1,0,1,2$, with
the same notation for the sum of $s$-labels.

The conformal weights and $U(1)$ charges of the highest weight states
of the coset representations are, up to integers
\beqa
h_{j,n,s} &=& h_j-h_n+h_s\ \ {\rm mod} \ 1  \\ &=&
\frac{j(j+1)}{k+2} -\frac{n^2}{4(k+2)} + \frac{s^2}{8} \ \ 
{\rm mod} \ 1 \\
q_{j,n,s} &=& \frac{s}{2} -\frac{n}{k+2} \ \ {\rm mod} \ 2
\eeqa
These relations hold exactly in the so-called standard range,
where $|n-s|\leq 2j$. In the context of orientifolds, one often
needs the square root of the $T$ transformation and therefore the
value of $h$ modulo $2$. In the coset model, one introduces the
sign factors $\sigma_{j,n,s}$ \cite{BH2}
\beq
\sigma_{j,n,s} = T^{\frac{1}{2}}_{[j,n,s]}T^{-\half}_j T^{\half}_n T^{-\half}_s,\eeq
where the $T$ matrices with single labels appearing 
on the right hand side are those of the $su(2)$
and $u(1)$ theories. The explicit value of $\sigma_{j,n,s}$ can be
calculated by studying the realization of the coset model ground states
within the representation spaces of $su(2)_k \oplus u(1)_2$, see the appendix
of \cite{BH2} for the explicit calculation. The result is
\begin{eqnarray}
\sigma_{j,n,s}=\left\{\begin{array}{ll} 1 &\mbox{ for } (j,n,s)\in \mbox{Standard range}\\
						1 & \mbox{ for } (j,-2j,2),2j\geq 1\\ 
						1 & \mbox{ for } (\frac{k}{2},k+2,0)\\
						(-1)^ {\frac{|n|-|s|}{2}-j} & \mbox{ for } (\frac{k}{2}-j,n \hat{+}(k+2),s\hat{+}2)\in \mbox{Std. range}\\
						-1 & \mbox{ for } (j,2j+2,0), j\neq \frac{k}{2}\\
						-1 & \mbox{ for } (j,\pm(2j+1), \mp 1)\\
						-1 & \mbox{ for } (0,0,2)\end{array}\right.
\end{eqnarray}
The Hilbert space for the coset model is
\beq\label{cosethilbert}
\H = \bigoplus_{[j,n,s]} \H_{[j,n,s]} \otimes \bar\H_{[j,n,\pm s]},
\eeq
where the choice of sign corresponds to two different GSO projections,
type 0A and type 0B. On the other hand, the representation spaces of 
the $N=2$ super-Virasoro
algebra (no GSO projection) 
are in terms of the representation spaces of the coset theory
\beq\label{N2hilbert}
\H^{N=2}_{[j,n,[s]]} = \H_{[j,n,s]} \oplus \H_{[j,n,s+2]} \ ,
\eeq
where $[s]$ is the $\ZZ_2$ reduction of the $\ZZ_4$ label $s$ and takes
the value $1$ for the Ramond and $0$ for the Neveu-Schwarz sector.
In the theory without GSO projection, the fields are  organized in
these representations of the $N=2$ algebra.

In this section, we are interested in the crosscap states in the
tensor product of two minimal $N=2$ models with a single GSO projection.
The Hilbert space of such a theory is
\ba\label{N2GSOhilbert}
\H & = & \bigoplus_{[j_1,n_1,s_1],[j_2,n_2,s_2]}
\Bigl( \left(\H_{[j_1,n_1,s_1]} \otimes \H_{[j_2,n_2,s_2]} \right) 
\otimes \left( \bar\H_{[j_1,n_1,s_1]} \otimes \bar\H_{[j_2,n_2,s_2]}
\right)  \\
& & \qquad  \qquad
\oplus \,\, 
\left(\H_{[j_1,n_1,s_1]} \otimes \H_{[j_2,n_2,s_2]} \right) 
\otimes \left( \bar\H_{[j_1,n_1,s_1+2]} \otimes 
\bar\H_{[j_2,n_2,s_2+2]} \right) \Bigr) \,, \nonumber
\ea
where the sums over $s_1$ and $s_2$ are restricted to 
$s_1-s_2\in 2\ZZ$. 
In comparison to the tensor product of two coset models, this
theory is spin aligned, so that the states of both factors are
either in the NSNS or RR sector. This projection gives rise to
a twisted sector, where the $s_i$ values for the left- and right
movers differ by $2$, corresponding to the lower line of (\ref{N2GSOhilbert}).

The formulas we have developed so far for permutation crosscap states
are directly applicable to
a tensor product of two coset models, where the Hilbert space is given by
the tensor product of two copies of (\ref{cosethilbert}). The
simple currents of a single copy of the coset theory carry representation
labels $(0,\nu,\sigma)$, and hence we can label the crosscap states by
$(\nu, \sigma)$. 
For the tensor product of two coset models, the crosscap state would thus read
\beq\label{tensorcross}
\vert \Scr{C}, \nu, \sigma \rangle_{(12)}^{coset} = \sum_{[j,n,s]} 
\frac{S_{[0,\nu,\sigma][j,n,s]}}{S_{[0,0,0][j,n,s]}} \cket{[j,n,s][j,\pm n, \pm s]}_{(12)} \ .
\eeq
The choice of sign in the Ishibashi state is just like in the $U(1)$ case and
refers to different boundary conditions, A-type $(+)$, and B-type $(-)$. We 
will specialize to the B-type case, the discussion for A-type permutation
crosscap states is completely
analogous. 

Note that the  crosscap state for the tensor product of
two coset models already obeys the spin alignment condition
$s_1-s_2 \in 2\ZZ$.
To understand how it has to be modified for the GSO projected
tensor product of two minimal $N=2$ models, 
we consider the Klein bottle amplitude calculated from
(\ref{tensorcross})
\beq\label{cosetkb}
{\cal K}_{(12)}^{coset}((\nu,\sigma),(\nu',\sigma')) = 
\sum_{[j_i,n_i,s_i]} 
\delta_{j_1j_2} \ \delta^{(2k+4)}_{n_1+\nu'-\nu,n_2} 
\delta^{(4)}_{s_1+\sigma'-\sigma,s_2} \ \chi_{[j_1,n_1,s_1]}(2\tau) \ 
\chi_{[j_2,n_2,s_2]} (2\tau) \ .
\eeq
This Klein bottle amplitude obviously only gets contributions from
the first part of the Hilbert space (\ref{N2GSOhilbert}). The orientifold
action in the closed string sector of this model can be read off to
be  (for $\nu'=\nu, \sigma'=\sigma$) a combination of a permutation
and a parity action in a single coset model. On the closed string
ground states it acts as
\beqa\label{parityuntwist}
&& (j_1,n_1,s_1)\otimes (j_2,n_2,s_2) \otimes \overline{(j_1,n_1,s_1)}
\otimes \overline{(j_2,n_2,s_2)} \to \\ \nonumber
&&~~~~~~~~~~~~~~~(j_2,n_2,s_2) \otimes (j_1,n_1,s_1)
\otimes \overline{(j_2,n_2,s_2)} \otimes \overline{(j_1,n_1,s_1)} \ .
\eeqa
The parity invariant states are those with $j_1=j_2$, $n_1=n_2$, $s_1=s_2$,
leading to the Klein bottle (\ref{cosetkb}). The natural extension
of this parity action to the twisted sector part is given by
\beqa\label{paritytwist}
&& (j_1,n_1,s_1)\otimes (j_2,n_2,s_2) \otimes \overline{(j_1,n_1,s_1+2)}
\otimes \overline{(j_2,n_2,s_2+2)} \to \\ \nonumber
&& ~~~~~~~~~~~~~~~(j_2,n_2,s_2+2) \otimes (j_1,n_1,s_1+2) \otimes
\overline{(j_2,n_2,s_2)} \otimes \overline{(j_1,n_1,s_1)} \ ,
\eeqa
such that state with $j_1=j_2$, $n_1=n_2$, $s_1=s_2+2$ are parity invariant.

The crosscap states that implement this type of parity action 
on the closed string sector states is given by projections of the
coset model crosscap states onto the NSNS (or RR) sector
\beqa 
\vert \Scr{C}, \nu, \sigma \rangle_{(12)}^{NSNS} &=& 
\sum_{[j,n,s]s \, {\rm ev}}
\frac{S_{[0,\nu,\sigma][j,n,s]}}{S_{[0,0,0][j,n,s]}} 
\cket{[j,n,s][j,-n,-s]}_{(12)} \\ \nn
\vert \Scr{C}, \nu, \sigma \rangle_{(12)}^{RR} &=& \sum_{[j,n,s]s \, {\rm odd}}
\frac{S_{[0,\nu,\sigma][j,n,s]}}{S_{[0,0,0][j,n,s]}} \cket{[j,n,s][j,-n,-s]}_{(12)} \ .
\eeqa
The corresponding Klein bottles are
\beqa
{\cal K}_{(12)}^{NSNS}  ((\nu,\sigma),(\nu',\sigma'))&=& 
\sum_{[j_i,n_i,s_i]} 
\delta_{j_1j_2} \ \delta^{(2k+4)}_{n_1+\nu'-\nu,n_2} \times \\ \nonumber
&&\big( \delta^{(4)}_{s_1+\sigma'-\sigma,s_2} + 
\delta^{(4)}_{s_1+\sigma'-\sigma+2,s_2}
\big)
\ \chi_{[j_1,n_1,s_1]}(2\tau) \ 
\chi_{[j_2,n_2,s_2]} (2\tau) \\ \nonumber
{\cal K}_{(12)}^{RR}  ((\nu,\sigma),(\nu',\sigma')) 
&=&\sum_{[j_i,n_i,s_i]} 
\delta_{j_1j_2} \ \delta^{(2k+4)}_{n_1+\nu'-\nu,n_2} \times \\
&& \big( \delta^{(4)}_{s_1+\sigma'-\sigma,s_2} -
\delta^{(4)}_{s_1+\sigma'-\sigma+2,s_2}
\big)
\ \chi_{[j_1,n_1,s_1]}(2\tau ) \ 
\chi_{[j_2,n_2,s_2]} (2 \tau )\ .
\eeqa
For $\sigma=\sigma'$, $\nu=\nu'$ the Klein bottle 
in the NSNS sector encodes the parity action 
(\ref{parityuntwist}) and (\ref{paritytwist})
in the closed string loop channel. The RR sector crosscap state leads
in this case to an additional insertion of $(-1)^{\frac{s_1-2_2}{2}}$,
which is the difference in fermion number in the two minimal model factors.
\footnote{Note that this type of  argument does not fix the phases appearing in front of
the crosscap Ishibashi states, since any such phase drops out from
our calculation.}

There is another obvious modification of the parity action, namely
we can dress it by space-time fermion number, which is $-1$ on the RR
sector states, and $+1$ on the NSNS sector states.
\beqa\label{parityuntwistmod}
&& (j_1,n_1,s_1)\otimes (j_2,n_2,s_2) \otimes \overline{(j_1,n_1,s_1)}
\otimes \overline{(j_2,n_2,s_2)} \to \\
&&~~~~~~~~~~~~~~~(-1)^{s_2} \ (j_2,n_2,s_2) \otimes (j_1,n_1,s_1)
\otimes \overline{(j_2,n_2,s_2)} \otimes \overline{(j_1,n_1,s_1)} \ 
\eeqa
and likewise its extension to the twisted sector. The crosscap states
leading to this modification contain only crosscap Ishibashi states
that are built on ground states coming from the twisted sector where
$s_i=\bar{s}_i+2$ for $i=1,2$. 
The corresponding crosscap states are
\beqa\label{twistedcrosscap} \nn
\!\!\!\!\!\!\!\vert \Scr{C}, \nu, \sigma \rangle_{(12)}^{NSNS} &=& 
\sum_{\substack{[j,n,s]\\s \, {\rm even}}} \ \sqrt{\frac{T_{j,-n,-(s\hat{+}2)}}{T_{j,-n,-s}}}
\frac{S_{[0,\nu,\sigma][j,n,s]}}{S_{[0,0,0][j,n,s]}} \cket{[j,n,s][j,-n,-s]}_{(12)} \\
\!\!\!\!\!\!\!\vert \Scr{C}, \nu, \sigma \rangle_{(12)}^{RR} &=& \sum_{\substack{[j,n,s]\\s \, {\rm odd}}}
\sqrt{\frac{T_{j,-n,-(s\hat{+}2)}}{T_{j,-n,-s}}} \
\frac{S_{[0,\nu,\sigma][j,n,s]}}{S_{[0,0,0][j,n,s]}} \cket{[j,n,s][j,-n,-s]}_{(12)} 
\eeqa
The $(jns)$-dependent phase factors introduced in front of the Ishibashi states
drop out from the Klein bottle calculation but will be motivated in
a moment.

Let us now turn to a discussion of the open string one loop amplitudes.
The B-type permutation branes are characterized by the gluing
conditions 
\ba
\left(L^{(1)}_n - \bar{L}^{(2)}_{-n} \right) 
| \Scr{B} \rangle\!\rangle =
\left(L^{(2)}_n - \bar{L}^{(1)}_{-n} \right) 
|\Scr{B} \rangle\!\rangle 
& = & 0
\nonumber \\
\left(J^{(1)}_n + \bar{J}^{(2)}_{-n} \right) |\Scr{B} \rangle\!\rangle 
= \left(J^{(2)}_n + \bar{J}^{(1)}_{-n} \right) |\Scr{B} \rangle\!\rangle 
& = & 0 
\label{gluingp} \\
\left(G^{\pm (1)}_r + i \eta_1 \bar{G}^{\pm (2)}_{-r} \right) 
|\Scr{B} \rangle\!\rangle = 
\left(G^{\pm (2)}_r + i \eta_2 \bar{G}^{\pm (1)}_{-r} \right) 
|\Scr{B} \rangle\!\rangle
& = & 0 \,. \nonumber
\ea
For A-type permutation branes, one would introduce a minus sign in the
condition for the $U(1)$ current $J$, and hence the conditions on the
supercharges would have to be modified accordingly.

The B-type permutation boundary states \cite{BG,ERR}  for the  GSO 
projected theory are given by
\beqa\label{permN2bs}
&& |\Scr{B}, J,N,S_1,S_2 \rangle_{(12)} = \frac{1}{\sqrt{2}} \sum_{[j,n,s]}
\frac{S_{[J,N,S_2-S_1][j,n,s]}}{S_{[0,0,0],[j,n,s]}} \ \times   \\ \nonumber
&& \times\big( \kket{([j,n,s][j,-n,-s])}_{(12)} +(-1)^{S_2}
  \kket{([j,n,s][j,-n,-(s\hat{+}2)])}_{(12)} \big) \ .
\eeqa
The labels $[J,N,S_1,S_2]$ are subject to the  
constraint $2J+N+S_1-S_2$ even. $S_1$ and $S_2$
are $\ZZ_4$ labels, and $S_1-S_2$ has to be even to preserve the diagonal $N=2$.
Compared to equation (\ref{gluingp}), $(-1)^{S_1}=\eta_1$ and 
$(-1)^{S_2}=\eta_2$. The boundary state is subject to the identification
$[J,N,S_1,S_2]= [J,N,S_1+2,S_2+2]$, and shifting either $S_1$ or $S_2$
by $2$ corresponds to mapping brane to anti-brane. The open string partition
function has been given in \cite{BG} and reads
\ba
&  & {\cal C}_{(12)(12)} ([J,N,S_1,S_2],[\hat{J},\hat{N},\hat{S}_1,\hat{S}_2])
(\tau)
=  \sum_{[j_i,n_i,s_i]}  
\chi_{[j_1,n_1,s_1]}(\tau)\, 
\chi_{[j_2,n_2,s_2]}(\tau)\times \nonumber
\\
& & \sum_{\hat{j}}
\Bigl[ N_{\hat{j} \hat{J}}{}^{J} \, N_{j_1 j_2}{}^{\hat{j}}\,
\delta^{(2k+4)}(\Delta N+n_1-n_2) \nonumber \\
& & \qquad \times
\Bigl( \delta^{(4)}(\Delta S_1+s_1)\, \delta^{(4)}(\Delta S_2+s_2)
+ \delta^{(4)}(\Delta S_1 +2+s_1)\, \delta^{(4)}(\Delta S_2 +2+s_2) 
\Bigr) \nonumber \\
& & \quad + N_{\hat{j}\, \frac{k}{2}-\hat{J}}{}^{J} \, N_{j_1 j_2}{}^{\hat{j}}\,
\delta^{(2k+4)}(\Delta N+k+2+n_1-n_2) \nonumber \\
& & \qquad \times 
\Bigl( \delta^{(4)}(\Delta S_1+2+s_1)\, \delta^{(4)}(\Delta S_2+s_2) 
+ \delta^{(4)}(\Delta S_1+s_1)\, \delta^{(4)}(\Delta S_2+2+s_2) 
\Bigr) \Bigr] \,, \nonumber
\ea
where $\Delta N = \hat{N} - N$ and $\Delta S_i = \hat{S}_i-S_i$.

Using the properties of the $P$ and $Y$ matrix of the minimal
model listed in the appendix, we now derive the following M\"obius
strip for the NSNS crosscap in the untwisted sector
\beqa\label{NSNSmoebius}
&& {\cal M}^{NSNS}_{(12)(12)}  ((\nu,\sigma) (J,N,S_1,S_2))(\tau) = \\ \nonumber&&=
\sum_{[j_i,n_i,s_i]} \sigma_{j_1,n_1,s_1} \sigma_{j_2,n_2,s_2} 
 \delta^{(2)}_{s_1} \ \delta^{(2)}_{s_2}\hat\chi_{[j_1,n_1,s_1]} (\tau) 
\hat\chi_{[j_2,n_2,s_2]} (\tau)  \times \ \\ \nn
 && \big(Y_{Jj_1}^{j_2} \ Y_{N-\nu \ n_1}^{n_2} \ 
\delta^{(4)}_{2(S_2-S_1-\sigma) + s_1-s_2}
 + Y_{J j_1}^{\frac{k}{2}-j_2}
Y_{N-\nu \ n_1}^{n_2 \hat{+}(k+2)} \delta^{(4)}_{2(S_2-S_1-\sigma) +s_1-s_2+2}
 (-1)^{\frac{|n|-|s|-2j}{2}} \big) 
\eeqa
One sees that for the $SU(2)$ and $U(1)_{k+2}$ part of the theory the 
orientifold action is inherited from that of the constituent theories.
In particular, the brane labels get mapped as
\beq
J \to J, \qquad N \to 2\nu-N \ .
\eeq
It remains to analyze the labels of the $U(1)_2$ part. By comparison
of the M\"obius strip with the cylinder amplitude one concludes that
\beq
\sigma=0: S_i \to S_i \ .
\eeq
The $S_i$ labels of the boundary state remain unaffected. On the
other hand, for the crosscaps with $\sigma=1$ one obtains
\beq
\sigma=1: S_1 \to S_1+2, \quad S_2 \to S_2,
\eeq
where equivalently one can exchange the roles of $S_1$  and $S_2$ by
the freedom to shift both boundary state labels by $2$. To summarize,
the crosscap with $\sigma=0$ maps branes to branes and anti-branes to
antibranes, the crosscap with $\sigma=1$ exchanges branes and anti-branes.
In a geometrical context, one would interpret such a brane-antibrane
flip as an orientation reversal of the cycle wrapped by the brane. 

For completeness, we also list the M\"obius amplitude for
the RR crosscap. The result is
\beqa
 &&{\cal M}^{RR}_{(12)(12)}  ((\nu,\sigma)(J,N,S_1,S_2))
= \sum_{j_i,n_i,s_i} \sigma_{j_1n_1s_1} \sigma_{j_2n_2s_2} \times \\ \nn
&& \big( \delta^{(4)}_{S_2-S_1-\sigma+\frac{s_1-s_2}{2}} 
- \delta^{(4)}_{S_2-S_1-\sigma+\frac{s_1-s_2}{2}+2} \big) \ \delta^{(2)}_{s_1+1} \ 
\delta^{(2)}_{s_2+1} Y_{Jj_1}^{j_2} Y_{N-\nu n_1}^{n_2}
\hat\chi_{(j_1,n_1,s_1)} (\tau) \hat\chi_{(j_2,n_2,s_2)} (\tau) \ .
\eeqa
In particular, the open string sector is in the R sector, which
means that the parity action exchanges even $S_i$ labels of the boundary
state with odd ones. A closer look at the amplitudes reveals that
\beqa
\sigma &=& 0: \quad S_1 \to S_1+1, \quad S_2\to S_2+1 \\
\sigma &=& 1: \quad S_1 \to S_1+1 \quad S_2 \to S_2-1 ,
\eeqa
so that we see the brane-antibrane flip also here.

Let us now turn to the amplitudes in the twisted sector.
Before going into the details, we can already predict what
the difference between the twisted and untwisted sector case
is: In the closed string sector, the two parity actions
differ by the operation that 
acts as $-1$ on the RR states and $+1$ on the NSNS. On the level
of boundary states, this means that the NSNS part of the boundary
state is unaffected, whereas the RR part is transformed with a $-$
sign. This means that the action of this parity on the branes
differs from the one discussed previously by
a brane-antibrane flip. So one would conclude that in the twisted NSNS sector
\beq\label{twistedstransformI}
\sigma=0: S_1 \to S_1+2, \quad S_2 \to S_2 \ ,
\eeq
whereas
\beq\label{twistedstransformII}
\sigma=1 : S_i \to S_i+2 \ .
\eeq
Similar arguments can be made for the RR sector.

Recall that $\nu$ has to be even in the case $\sigma=0$ and odd in the case
$\sigma=1$, such that there is a difference in the possible parity actions
between the untwisted and
twisted case.

Let us outline briefly how to obtain these results technically. First,
we need to motivate the extra phases appearing in the crosscap
state in the twisted sector. For this note that in the fully 
supersymmetric theory (where the Hilbert space is given by (\ref{N2hilbert}))
Ishibashi states from the twisted and untwisted sector have
to be combined to supersymmetric Ishibashi states. For the boundary
state, the supersymmetric Ishibashi states can be directly read
off from the boundary state (\ref{permN2bs}), explicitly
\beqa
&& \kket{([j,n,s][j,-n,-s])}^{N=2}_{(12)} = \Big(
\kket{[j,n,s][j,-n,-s]}_{(12)}  \\ \nonumber
&&+(-1)^{S_2}
\kket{[j,n,s][j,-n,-(s\hat{+}2)]}_{(12)} 
+(-1)^{S_1} \kket{[j,n,s\hat{+}2][j,-n,-s]}_{(12)}  \\ \nonumber
&&+(-1)^{S_1+S_2}\kket{[j,n,s\hat{+}2][j,-n,-(s\hat{+}2)]}_{(12)}  \Big),
\eeqa
The conditions to be fulfilled by 
the crosscap states can be obtained from those for
the boundary states (\ref{gluingp}) by conjugation with $e^{\pi i L_0}$,
where $L_0=L_0^{(1)} + L_0^{(2)}$. This conjugate condition is
fulfilled by 
\beq
e^{\pi i L_0} |{\Scr{B}} \rangle\!\rangle =
e^{\pi i h} |{\Scr C} \rangle\!\rangle 
\eeq
Applying $e^{\pi i L_0}$  to the supersymmetric boundary Ishibashi state
shows that for the $N=2$ supersymmetric crosscap Ishibashi state there
has to be a relative phase equal to $\sqrt{T_{[j,-n,-(s\hat{+}2)]} T^{-1}_{[j,-n,-s]}}$
between the crosscap Ishibashi states $\cket{[j,n,s][j,-n,-s]}$ and
$\cket{[j,n,s][j,-n,-(s\hat{+}2)]}$, as we have anticipated in 
(\ref{twistedcrosscap}). In the calculation of the M\"obius strip,
one makes use of the following symmetry property of the $P$-matrix
of the $N=2$ minimal model
\beqa
P_{[j,n,s][j^{\prime},n^{\prime},s^{\prime}+2]}\frac{\sqrt{T_{[j^{\prime},n^{\prime},s^{\prime}]}}}{\sqrt{T_{[j^{\prime},n^{\prime},s^{\prime}+2]}}}=\frac{\sqrt{T_{[j,n,s]}}}{\sqrt{T_{[j,n,s+2]}}}P_{[j,n,s+2][j^{\prime},n^{\prime},s^{\prime}]} \ .
\eeqa
Effectively, this shifts the $\delta^{(4)}$ functions for the $s$ labels
in the M\"obius strip (\ref{NSNSmoebius}) 
by $2$ verifying the action (\ref{twistedstransformI}),
(\ref{twistedstransformII}) .

\medskip
\noindent
We refer to \cite{hosomichi} for further discussions of the $N=2$
minimal models and their orbifolds. In addition, applications to Gepner models
and the construction of string vacua can be found in that paper.

\subsection{The Landau Ginzburg approach} \label{SUSYLG}

A Landau Ginzburg model of chiral superfields $X_i$ has the following action
\beq
S = \int d^4 \theta K(X_i,\bar{X}_i) + \int d\theta^-d\theta^+ W(X_i) 
\vert_{\bar\theta^{\pm}=0} + \int d\bar{\theta}^+d\bar{\theta}^- \bar{W}
(\bar{X}_i) 
\vert_{\theta^{\pm}=0} 
\eeq
Here, $\pm$ distinguishes left and right movers, whereas bar is the
complex conjugation. 
Parity symmetries of Landau Ginzburg theories have
been discussed in \cite{BH2,HWLG}. In particular, B-type parity acts on the
superspace coordinates as $\theta^\pm \to \theta^{\mp}$, 
$\bar\theta^\pm \to \bar\theta^{\mp}$. Obviously, the measure in the
F-term of the action picks up a sign under B-parity. 
For the action to be invariant, one must define an action on the superfields,
such that $W$ flips sign under the induced action

For the case of a single minimal model, the superpotential is
\beq
W=X^{k+2}
\eeq
and the action $X\mapsto -X$ leads to a parity invariant F-term in the case
that $k$ is odd. For the case $k$ even, there is no pure involutive 
B-type parity. 

For the tensor product theory, the relevant superpotential is 
\beq
W=X_1^{k+2} + X_2^{k+2}
\eeq
and a suitable involution is
\beq\label{LGinv}
\tau: X_1 \mapsto \eta X_2, \quad \quad X_2 \mapsto \eta^{-1} X_1,
\eeq
where $\eta$ is a $(k+2)$th root of $-1$, $ \eta^{k+2}=-1$. So for the
tensor product theory, there exists a B-parity for any $k$, which is what
we would have expected from our conformal field theory analysis, where
we have explicitly constructed B-crosscap states for any level.
Furthermore, it is suggestive that the choice of $\eta$ corresponds to
the choice of the crosscap state label $\nu$ in conformal field theory.

That this is indeed the case can be understood best 
in the language of matrix factorizations \cite{HWLG}.
Recall that B-type D-branes in Landau Ginzburg models are described
by matrix factorizations of the superpotential \cite{MK,DO,KL,BHLS,HLL}
\beq
W=E(X_i) F(X_i) ,
\eeq
where $E,F$ are matrices with polynomial entries.
The two polynomial matrices are arranged into a larger
matrix that plays the role
of a BRST operator for the theory with boundary
\beq
Q= \left( \begin{array}{cc} 0 & E \\ F & 0 \end{array} \right) \ .
\eeq
Two factorizations are equivalent, if they are related by a similarity
transformation
\beq
\hat{Q} = U Q U^{-1} \ ,
\eeq
where both $U$ and $U^{-1}$ are block diagonal matrices with polynomial
entries.
In our case, the
superpotential can be written as
\beq\label{permutationfactorization}
W=X_1^{k+2} + X_2^{k+2} = \prod_{\eta} (X_1-\eta X_2),
\eeq
where $\eta$ runs over all $(k+2)$th roots of $-1$. Suitable factorization
are then obtained by organizing the linear factors into two groups.
These factorizations and their geometric relevance have been studied
in \cite{ADD,ADDF}.
In \cite{BG,ERR,Gxx} a subclass of this type was identified
with permutation D-branes, namely
\beq\label{match}
|\Scr{B},J,N,S_1=S_2=0 \rangle \Leftrightarrow F=\prod_{m=(N-2J)/2)}^{(N+2J)/2}
(X_1-\eta_m X_2),
\eeq
where $\eta_m=e^{-\frac{\pi i (2m+1)}{k+2}}$, and the spin structures were
taken to be fixed, such that $2J+N$ even. It remains to be clarified
how to obtain conformal field theory descriptions for the other
factorizations obtained by grouping the factors in 
(\ref{permutationfactorization}) in arbitrary
ways.

In \cite{HWLG} it was realized that also orientifolds have a description
in terms of matrix factorizations. Namely, they are described by
the matrix factorization that corresponds to the D-brane localized on
the fixed point set. The (topological) crosscap state is the same as
the boundary state. In our case, the fixed point set of the involution
(\ref{LGinv}) is given by
\beq
X_1-\eta X_2 =0 \ ,
\eeq
to which one associates the matrix factorization with
\beq
F=X_1-\eta X_2
\eeq
According to the identification of boundary states with matrix factorizations,
this brane has boundary state label $J=0$ and $N$ (which has to be even)
is determined by the
phase $\eta(N)=\exp(\pi i (N+1)/(k+2))$. 
It is therefore suggestive that one of the 
CFT crosscap states labelled $\nu$ with
$\nu$ even corresponds on the LG side to the involution (\ref{LGinv})  
\beq\label{etaparametrize}
\eta=\eta(\nu)=e^{-\frac{\pi i (\nu +1)}{k+2}} \ .
\eeq
To see this more precisely, we will consider 
the action of the parity on the D-branes
in Landau Ginzburg language. 
In \cite{HWLG} (drawing from the unpublished work \cite{GH}) it was shown that the parity
induces the following operation on the boundary BRST operators
\beq
Q(X_i) \to -Q(\tau (X_i))^T ,
\eeq
where $(-)^T$ is the graded transpose
\beq
\left( \begin{array}{cc} a & b \\ c & d \end{array} \right)^T
= \left( \begin{array}{cc} a^t & -c^t \\ b^t & d^t  \end{array} \right) \ ,
\eeq
where $(-)^t$ denotes the ordinary transpose. 
The action is indeed quite natural \cite{HWLG}: If one associates
Chan Paton spaces $V$ and $W$ to the left and right boundary of an open string, 
the space of open string states includes the Chan Paton factor
\beq
\H^{{\rm CP}} = \Hom (V,W)
\eeq
A parity action exchanges the left and right boundary of the open string.
If left and right boundary were initially oppositely oriented, then the
new (after application of the parity) 
left  boundary is oppositely oriented compared to the
initial left boundary. The space of Chan-Paton factors is mapped to
$\Hom (W^*,V^*)$ which is naturally implemented on the open string
states by taking the transpose. In the case that the vector spaces
are graded, as is the case for Landau Ginzburg models, it is suggestive
that a graded version of the transpose should be used. This was
indeed confirmed in \cite{HWLG} by analyzing the parity action
with the help of the action for the boundary fermions appearing in
the Landau Ginzburg action.

This parity acts on the factors $E(X_i)$ and $F(X_i)$ as   
$F_{image}= - \tau^* E$, and $E_{image}=\tau^*F$, such
that $F_{image} E_{image}=-W$ is a factorization of the parity-transformed
superpotential.
For the factorization with $F$ as in (\ref{match}) this means
that the image factorization under the parity (\ref{LGinv}) is given as
\beq 
E_{image}= 
-\prod_{n=(N-2J)/2)}^{(N+2J)/2} (\eta X_2 - \eta_n \eta(\nu)^{-1} X_1)
= -\prod_{n=(N-2J)/2)}^{(N+2J)/2} (-\eta^{-1}\eta_n) \ ( X_1 - 
\eta^2\eta_n^{-1} X_2) \ .
\eeq
Parametrizing the possible $\eta$ by even integers $\nu$ as in
(\ref{etaparametrize}) allows us to write the image brane as
\beq
E_{image}  = \mbox{const}\times \prod_{n=(2\nu-N+2J)/2}^{(2\nu-N-2J)/2} (X_1-\eta_n X_2) \ .
\eeq
This is, up to an exchange of the factors $E$ and $F$ which corresponds
to an orientation flip, indeed the factorization associated to the
boundary state with $N\to 2\nu-N$, $J\to J$. 
The action $N \to 2\nu-N$ and $J\to J$ is indeed the parity action that
we also observed in conformal field theory. Note however that in conformal
field theory, depending on the value of the crosscap label $\sigma$, $\nu$
could take even or odd labels. In the Landau Ginzburg theory, the $\nu$
labels are even (with the choices we have made), in particular, the
reflection $N \to -N$ is a possible parity action on the brane labels.
The natural action on the Chan-Paton factors by a graded transpose
induces an orientation flip of the Landau-Ginzburg brane. In the conformal
field theory, there was a choice of parities, one of them leading to
a brane-anti-brane flip, the other not. Here we have seen that the parity
with flip corresponds directly to the Landau Ginzburg parity. To conclude,
the parity action on brane labels in (\ref{twistedstransformI})
together with $J\to J$ and $N \to 2\nu-N$, $\nu$ even is the CFT
description (in the NSNS sector) of the Landau Ginzburg parity.

\medskip
\medskip

\centerline {\large \bf Acknowledgements}  
We would like to thank Stefan Fredenhagen,
Matthias Gaberdiel and Ingo Runkel 
for discussions. We thank Kazuo Hosomichi for informing us
about his work \cite{hosomichi} .
This work was supported by the
EURYI award of the European Science Foundation,
a TH-grant from ETH Z\"urich, and the Marie-Curie network Forces Universe
(MRTN-CT-2004-005104).

\newpage

\appendix 
\section{Loop amplitudes}

First, we define the following set of matrices from the fusion and $Y$ coefficients
\begin{equation}
\label{Y definition}
(N_i)_{mn}:=N_{im}^n=\sum_j\frac{S_{ij}S_{mj}\bar{S}_{nj}}{S_{0j}}\quad \mbox{and }\quad(Y_i)_{mn}:=Y_{im}^n=\sum_j\frac{S_{ij}P_{mj}\bar{P}_{nj}}{S_{0j}}
\end{equation}
where $S$ are the usual modular matrices and $P:=\sqrt{T}ST^2S\sqrt{T}$.
Tensor boundary and crosscap states are denoted  with an additional
subscript $1$ and carry a 
boundary index $\alpha$ composed of two labels $\alpha=(\alpha_1,\alpha_2)$, 
whereas the permutation states are denoted with the transposition 
$(1\ 2)$ and with only one boundary label $\alpha$. 
In the following, $\alpha$ and $\beta$  are boundary state labels, 
whereas $\mu$ and $\nu$ refer to  crosscaps, and are in particular
simple currents.
Then, the one loop amplitudes for the conjugate modular invariant are expressed, first for the cylinder case, as
\begin{eqnarray}
\mathcal{C}_{11}(\alpha,\beta)&=&\sum_{i}\chi_i(\tau)(N_i)_{\beta_1 \alpha_1}\sum_j\chi_j(\tau)(N_j)_{\beta_2\alpha_2}\nonumber\\
\mathcal{C}_{1(1\ 2)}(\alpha,\beta)&=& \sum_{i}\chi_i\left(\frac{\tau}{2}\right)(N_iN_{\bar{\alpha}_1})_{\beta\alpha_2}\nonumber\\
\mathcal{C}_{(1\ 2)(1\ 2)}(\alpha,\beta)&=& \sum_{i,j}\chi_i(\tau)\chi_j(\tau)(N_i N_j)_{\beta\alpha} 
\end{eqnarray}
As for the Klein bottle amplitudes, the results are
\begin{eqnarray}
\mathcal{K}_{11}(\mu,\nu)&=&\sum_i\chi_i(2\tau)(Y_i)_{\nu_1\mu_1}\sum_j\chi_j(2\tau)(Y_j)_{\nu_2\mu_2}\nonumber\\
	\mathcal{K}_{1(1 \ 2)}(\mu,\nu)&=&\sum_i\chi_i(\tau)(Y_iY_{\nu})_{\bar{\mu}_1\mu_2}\nonumber\\
	\mathcal{K}_{(1\ 2)(1 \ 2)}(\mu,\nu)&=&\sum_{i,j}\chi_i(2\tau)\chi_j(2\tau)(N_jN_i)_{\nu\mu}
\end{eqnarray}
Finally, we obtain for the M\"{o}bius strip case the results
\begin{eqnarray}
\mathcal{M}_{1 1}(\mu,\alpha)&=&\sum_i\hat{\chi}_i(\tau)(Y_{\alpha_1})_{i\mu_1}\sum_j\hat{\chi}_j(\tau)(Y_{\alpha_2})_{j\mu_2}\nonumber\\
	\mathcal{M}_{(1 \ 2) 1}(\mu,\alpha)&=&\sum_i\hat{\chi}_i(2\tau)(N_{\alpha_1}N_{\alpha_2})_{i\mu}\nonumber\\
	\mathcal{M}_{1 (1 \ 2)}(\mu,\alpha)&=&\sum_i\hat{\chi}_i(2\tau)(Y_{\alpha}Y_i)_{\bar{\mu}_1\mu_2}\nonumber\\
	\mathcal{M}_{(1 \ 2) (1 \ 2)}(\mu,\alpha)&=&\sum_{i,j}\hat{\chi}_i(\tau)\hat{\chi}_j(\tau)(Y_{\alpha} Y_{\bar{\mu}})_{i\bar{j}},
\end{eqnarray}
where we remark that in our convention $\mathcal{M}=\langle\Scr{C}|e^{-\frac{\pi i H_c}{4\tau}}|\Scr{B}\rangle$.

\section{ The $Y$ tensor of SU(2) } \label{Ytensor}
The coefficients of the $Y$ tensor in the case of $\widehat{su}_k(2)$ are given by
\beqa
Y_{ij}^n=\delta^{(2)}_{2i}\delta_{jn}&+&\sum_{l=1,2i+l\in 2\mathbb{N}}^{2i}\Big[\delta_{j+l,n}+\delta_{j,n+l}-\delta_{j+n+1,l}\nonumber\\&&-(-1)^{2j+k}(\delta_{j+l,n+k+2}+\delta_{j+k+2,n+l}-\delta_{j+n+l,k+1})\Big]
\eeqa
In particular, these formulas imply that $Y_{ij}^n$ is zero whenever $j\in \mathbb{N}$ and $n\in \mathbb{N}+\frac{1}{2}$ or $j\in \mathbb{N}+\frac{1}{2}$ and $n\in \mathbb{N}$. This means that $(-1)^{2j}=(-1)^{2n}$ so the $Y_i$ matrices are symmetric in $j,n$ as they should be.\\
Of great utility to us are going to be the following simpler expressions
\begin{eqnarray}
	&&Y_{i0}^0=(-1)^{2i}\qquad Y_{i\frac{k}{2}}^{\frac{k}{2}}=1\ \forall i\qquad Y_{i\frac{k}{2}}^0=\frac{1+(-1)^k}{2}\delta_{i\frac{k}{4}}=\delta^{(2)}_k\delta_{i\frac{k}{4}}\nonumber\\ && Y_{\frac{k}{4}i}^{\frac{k}{2}}=\delta^{(2)}_k\delta^{(2)}_{2i}\qquad Y_{\frac{k}{4}i}^0=\delta^{(2)}_k\delta^{(2)}_{2i}(-1)^{i+\frac{k}{2}}
 \end{eqnarray}
Similarly, the $Y$ tensor has the symmetry relations
 \begin{eqnarray}
 &&Y_{ij}^n=Y_{in}^j\qquad Y_{i\frac{k}{2}-j}^{\frac{k}{2}-n}=(-1)^{2i+j-n}Y_{ij}^n
 \nonumber\\ &&Y_{\frac{k}{2}-i\ j}^n=(-1)^{k+2j}Y_{ij}^n\qquad Y_{i\frac{k}{2}-j}^n=(-1)^{2i+\frac{k}{2}+j+n}Y_{ij}^{\frac{k}{2}-n}
\end{eqnarray}
All of these formulas can be easily derived from the general one, or from the symmetries of the SU(2) $P$ matrix. \\
In addition to all this, there is a formula that relates the fusion coefficients to the $Y$ tensor, namely
\begin{equation}
Y_{Jj_1}^{j_2}=Z_{Jj_1}^{j_2}\sum_{j=0}^{[\frac{k}{2}]}N_{JJ}^jN_{j_1j_2}^j(-1)^j
\end{equation}
where $Z_{Jj_1}^{j_2}$ is only a sign. Since the sum on the right hand side is invariant under $J\mapsto \frac{k}{2}-J$ or under $(j_1,j_2)\mapsto (\frac{k}{2}-j_1,\frac{k}{2}-j_2)$, we need only to give the values of $Z_{Jj_1}^{j_2}$ for $J\leq \frac{k}{4}$ and $j_1\leq \frac{k}{4}$, because the rest can be found by using the symmetry relations for the $Y$ tensor. One finds
\begin{eqnarray}
Z_{Jj_1}^{j_2}=(-1)^{2J}\epsilon\qquad \mbox{ for } J,j_1\leq \frac{k}{4}
\end{eqnarray}
where $\epsilon$ is equal to one except when all of the following conditions are fulfilled:  $j_2>\frac{k}{4}$, $2J\geq\min\{j_1+j_2,k-j_1-j_2\}$, $2J+j_2-j_1$ is odd and one of the following is true:
\begin{itemize}
\item $k$ is odd and $j_1\in \mathbb{N}+\frac{1}{2}$ 
\item $k$ is even, $j_1\in \mathbb{N}$ and $j_1+j_2>\frac{k}{2}$.
\end{itemize}
Putting all together, we write
\begin{eqnarray}
&&Z_{Jj_1}^{j_2}=(-1)^{2J}\epsilon\qquad Z_{\frac{k}{2}-Jj_1}^{j_2}=(-1)^{2J+2j_1+k}\epsilon\nonumber\\
&&Z_{J\frac{k}{2}-j_1}^{\frac{k}{2}-j_2}=(-1)^{j_1-j_2}\epsilon\qquad Z_{\frac{k}{2}-J\frac{k}{2}-j_1}^{\frac{k}{2}-j_2}=(-1)^{2J+j_1+j_2+k}\epsilon
\end{eqnarray}
with $\epsilon$ as above and $J,j_1\leq \frac{k}{4}$.
\medskip
\medskip

\section{$P$ and $Y$ matrices}

For convenience, we list some orientifold specific data for the
theories $SU(2)_k$, $U(1)_k$ and the coset model, all of which
can be found in the appendix of \cite{BH2}.

\subsubsection*{\underline{$SU(2)_k$}}

The P-matrix of the level k $SU(2)$ WZW model is
$$
P_{j j'}=\frac{2}{\sqrt{k+2}}
\sin\left[\frac{\pi(2j+1)(2j'+1)}{2(k+2)}\right]\delta_{2j+2j'+k}^{(2)}.
$$
The components of the $Y$ tensor are explicitly evaluated in the
separate appendix \ref{Ytensor}.

\subsubsection*{\underline{$U(1)_k$}}

The P-matrix and Y-tensor of the level $k$ $U(1)$ is
\beqa
&P_{n n'}=\frac{1}{\sqrt{k}}\e^{-\frac{\pi i \widehat{n}\widehat{n'}}{2k}}
\delta_{n+n'+k}^{(2)},
\nn\\[0.2cm]
&Y_{n n'}^{n''}
=\delta_{n'+n''}^{(2)}\left(
\delta_{n+\frac{\widehat{n'}-\widehat{n''}}{2}}^{(2k)}
+(-1)^{n'+k}
\delta_{n+\frac{\widehat{n'}-\widehat{n''}}{2}+k}^{(2k)}\right),
\nn
\eeqa
where $\widehat{n}$ is the unique member of $n+2k\ZZ$ in the standard range
$\{-k+1,...,k-1,k\}$. In the following, we will omit the $\hat{}$, but it
is understood that all labels are chosen in this way.

\subsubsection*{\it \underline{Minimal model}}

We first note that the Q-matrix $Q=ST^2S$ of the minimal model
can be expressed in terms of the Q-matrices of the constituent
theories in the following way
\beq
Q_{(j,n,s)(j',n',s')} = Q_{jj'} Q_{nn'}^* Q_{ss'} + 
Q_{j(\frac{k}{2}-j')} Q_{n(n'\hat{+}(k+2))}^* Q_{s(s'\hat{+}2)}
\label{factorizedformofQ}
\eeq
The P-matrix is then obtained as
\beqa
P_{(j,n,s)(j'n's')} &=& T_{(j,n,s)}^{\frac{1}{2}} Q_{(j,n,s)(j',n',s')}
T_{(j',n',s')}^{\frac{1}{2}} \\ \no
&=& \sigma_{j,n,s} \sigma_{j'n's'} P_{jj'} P_{nn'}^* P_{ss'} \\ \no
&&~~~+ \sigma_{j,n,s} \sigma_{\frac{k}{2}-j', n'\hat{+}(k+2), s'\hat{+} 2}
P_{j,\frac{k}{2}-j'} P_{n,n'\hat{+}(k+2)}^* P_{s,s'\hat{+}2} \\ \no
&=&\sigma_{j,n,s} \sigma_{j'n's'} \left( P_{jj'} P_{nn'}^* P_{ss'}
+(-1)^{\frac{|n'|-|s'|-2j'}{2}}
P_{j,\frac{k}{2}-j'} P_{n,n'\hat{+}(k+2)}^* P_{s,s'\hat{+}2} \right)
\eeqa
\noindent
An explicit expression for the P-matrix is
\beqa\no
P_{(j,n,s)(j'n's')}\!\!&=&\!\!
 \sigma_{j,n,s} \sigma_{j'n's'} \frac{\sqrt{2}}{k+2}
\delta^{(2)}_{s+s'} \ \e^{\frac{\pi i nn'}{2(k+2)}} \ \e^{-\frac{\pi i ss'}{4}}
\Bigl( \sin\left[\pi\mbox{$\frac{(2j+1)(2j'+1)}{2(k+2)}$}\right]
 \delta^{(2)}_{2j+2j'+k}
\ \delta^{(2)}_{n+n'+k} \\ 
&& + (-1)^{\frac{2j'+n'+s'}{2}} \e^{\frac{\pi i (n+s)}{2}}
\sin\left[\pi \mbox{$\frac{(2j+1)(k-2j'+1)}{2(k+2)}$}\right]
 \delta^{(2)}_{2j+2j'}
\ \delta^{(2)}_{n+n'} \Bigr)
\eeqa
To decompose the $Y$ matrix for the minimal model into $Y$ matrices
of the constituent theories, one first introduces the quantity
\beq
\tilde{Y}_{ab}^c = \sum_d \frac{S_{ab} Q_{bd}Q^*_{cd}}{S_{0d}}
=\sqrt{\frac{T_c}{T_b}} \ Y_{ab}^c.
\eeq
One then finds
\beq
\tilde{Y}_{(j,n,s)(j',n',s')}^{(j'',n'',s'')} = \tilde{Y}_{jj'}^{j''} \
\overline{\tilde{Y}}_{nn'}^{n''} \tilde{Y}_{ss'}^{s''}+ 
\tilde{Y}_{jj'}^{\frac{k}{2}-j''} \
\overline{\tilde{Y}}_{nn'}^{n''+k+2} \tilde{Y}_{ss'}^{s''+2}.
\eeq
This decomposition allows us to evaluate the following combination
of $Y$-matrices
\beqa
&& Y_{(j,n,s)(j',n',s')}^{(j'',n'',s'')} + 
Y_{(j,n,s+2)(j',n',s')}^{(j'',n'',s'')} 
= 2\sigma_{j_1,n_1,s_1} \sigma_{j_2,n_2,s_2} 
\delta^{(2)}_{s'} \ \delta^{(2)}_{s''} \ \\ \nn
&& ~~~\big( Y_{jj'}^{j''} Y_{nn'}^{n''} \ \delta^{(4)}_{2s+s'-s''}
+ (-1)^{\frac{|n|-|s|-2j}{2}}
Y_{jj'}^{\frac{k}{2} -j''} Y_{nn'}^{n''\hat{+}(k+2)} \ 
\delta^{(4)}_{2s+s'-s''+2}
\big)
\eeqa
Here, we have made use of the explicit form of the $U(1)_2$ $Y$ matrix,
from which one can derive the property
\beq
Y_{ss'}^{s''} + Y_{s\hat{+} 2 \ s'}^{s''} = 2 \delta^{(2)}_{s'} \
\delta^{(2)}_{s''} \ \delta^{(4)}_{2s+s'-s''}
\eeq

\end{document}